\documentclass[a4paper,12pt]{article}

\usepackage[dvipdfmx]{graphicx}
\usepackage{amsmath}
\usepackage{amssymb}
\usepackage{amsfonts}
\usepackage{amsthm}
\usepackage{braket}
\usepackage{stmaryrd} % for \llbracket, \rrbracket
\usepackage{cite}

\allowdisplaybreaks

 \makeatletter
 \@addtoreset{equation}{section}
 \makeatother

\newcommand{\del}{\partial}

\newcommand{\bbra}[1]{\boldsymbol{(}#1\boldsymbol{)}} %( , )
\newcommand{\dop}{\mathrm{d}} %% differential operator
\newcommand{\iop}{\iota} %% interior product operator

\usepackage{color}
\usepackage[normalem]{ulem} % for cancel line
\usepackage{cancel} % for cancel

\renewcommand\sout{\bgroup \color{red} \ULdepth=-.5ex \ULset}

\date{empty}
\begin{document}
\begin{titlepage}
\null
\begin{flushright}
February, 2024
\end{flushright}
\vskip 2cm
\begin{center}
  {\Large \bf 
Extended Doubled Structures of Algebroids for
\\
\vspace{0.4cm}
Gauged Double Field Theory
}
\vskip 1.5cm
\normalsize
\renewcommand\thefootnote{\alph{footnote}}

{\large
Haruka Mori\footnote{h.mori(at)sci.kitasato-u.ac.jp} and 
Shin Sasaki\footnote{shin-s(at)kitasato-u.ac.jp}
}
\vskip 0.7cm
  {\it
  Department of Physics,  Kitasato University \\
  Sagamihara 252-0373, Japan
}
\vskip 2cm
\begin{abstract}
We study an analogue of the Drinfel'd double for algebroids associated with
 the $O(D,D+n)$ gauged double field theory (DFT).
We show that algebroids defined by the twisted C-bracket in the gauged DFT
are built out of a direct sum of three (twisted) Lie algebroids.
They exhibit a ``tripled'',  which we call the extended double, rather
 than the ``doubled'' structure appearing in (ungauged) DFT.
We find that the compatibilities of the extended doubled structure
 result not only in the strong constraint but also the additional condition in the gauged DFT.
We establish a geometrical implementation of these structures in a
 $(2D+n)$-dimensional product manifold and examine the relations to the generalized geometry
 for heterotic string theories and non-Abelian gauge symmetries in DFT.
\end{abstract}
\end{center}

\end{titlepage}

\newpage
\setcounter{footnote}{0}
\renewcommand\thefootnote{\arabic{footnote}}
\pagenumbering{arabic}
\tableofcontents
%%%%%%%%%%%%%%%%%%%%%%%%%%%%%%%%%%%%%%%%%%%%%%%%%
\section{Introduction} 
\label{sect:introduction}

Double field theory (DFT) is a field theory framework that encompasses
T-duality in string theories \cite{Hull:2009mi, Hohm:2010pp, Siegel:1993th}.
The theory is defined in the $2D$-dimensional doubled space
$\mathcal{M}_{2D}$ on which
fundamental fields $\mathcal{H}_{MN} (\mathbb{X}), \, (M,N = 1, \ldots, 2D)$ and $d (\mathbb{X})$ are defined.
Here $\mathcal{H}_{MN} (\mathbb{X})$ and $d (\mathbb{X})$ are the
generalized metric and the generalized dilaton, respectively.
The coordinate of the doubled space $\mathcal{M}_{2D}$ is decomposed as
$
\mathbb{X}^M = (\tilde{x}_{\mu}, x^{\mu}),
 \, (\mu = 1, \ldots, D)
$
 where 
$\tilde{x}_{\mu}$
 and 
$x^{\mu}$
may be identified with the winding and the Kaluza-Klein coordinates.
The theory exhibits manifest $O(D,D)$ structure and possesses a gauge symmetry.
In order to be physically consistent, all the quantities including fields and
gauge parameters in DFT must satisfy a physical condition called the strong constraint.
This indeed guarantees the gauge invariance of the action and closure of
the gauge algebra.
The strong constraint is trivially solved by quantities that depend only
on $x^{\mu}$. In this case, the action of DFT reduces to that of the
NSNS sector in type II supergravities.
A remarkable fact about DFT is that the diffeomorphism and the gauge
symmetry of the $B$-field are unified in a T-duality covariant way.
This is realized by the gauge symmetry in DFT whose algebra is governed by the C-bracket.

The algebraic structures defined by the C-bracket are known as
algebroids.
They are in fact a unification of geometries and symmetries.
The C-bracket is quite different from the Lie bracket.
Most notably, the C-bracket does not satisfy the Jacobi identity in general.
This leads to the notion of the metric (DFT or Vaisman) algebroid
\cite{Vaisman:2012ke, Chatzistavrakidis:2018ztm}.
For the $O(D,D)$ DFT, several algebroid structures have been studied
\cite{Mori:2019slw, Carow-Watamura:2020xij, Mori:2020yih,
Grewcoe:2020hyo, Marotta:2021sia, Marotta:2022tfe, Carow-Watamura:2022ten}.
They are strongly tied with geometric structures of $\mathcal{M}_{2D}$.
It is discussed that the doubled space is realized by a para-Hermitian geometry in which the metric
algebroid was first defined \cite{Vaisman:2012px}. The relations among the strong constraint,
physical spacetime and the Courant algebroid in the generalized geometry
have been studied \cite{Freidel:2017yuv, Svoboda:2018rci, Freidel:2018tkj, Marotta:2018myj}.
Among other things, {\it doubled structures} of algebroids play important
roles in DFT.
It is proven that the metric algebroid inherits a doubled structure \cite{Mori:2019slw}.
This is an analogue of the Drinfel'd double for Courant algebroids \cite{LiWeXu}
(see for the Drinfel'd double for Lie algebras \cite{Drinfeld:1983ky}).
The Drinfel'd double is based on a pair of the dual (maximally
isotropic) Lie algebroids $(L, \bar{L})$ by which a Manin triple 
is defined.
Indeed, upon imposing the strong constraint, the C-bracket becomes 
the Courant bracket and then the metric algebroid reduces to the Courant
algebroid \cite{Mori:2019slw, Hull:2009zb}.
The notion of the Drinfel'd double is necessary especially in the context of
the Poisson-Lie (and non-Abelian) T-duality (and plurality)
\cite{delaOssa:1992vci, Klimcik:1995ux, VonUnge:2002xjf}.
The duality (plurality) is interpreted as the exchange of the Manin triples for
given algebraic structures and it is a generalization of T-duality.
In this sense, the Poisson-Lie T-duality (U-duality) is naturally
understood in DFT and exceptional field theory (EFT)
\cite{Carow-Watamura:2022ten, Hassler:2017yza, Sakatani:2019jgu, Malek:2019xrf, Sakatani:2020wah, Musaev:2020nrt, Fernandez-Melgarejo:2021zyj, Bugden:2021nwl, Bugden:2021wxg}.

On the other hand, for heterotic string theories and gauged
supergravities, an alternative formalism of manifest T-duality known as
the gauged DFT has been developed \cite{Hohm:2011ex, Grana:2012rr}.
In the gauged DFT, non-Abelian gauge symmetries are introduced by gauging a
subalgebra of the extended duality group $O(D,D+n)$.
The gauge symmetry in DFT is modified and governed by the twisted
C-bracket. The physical condition is then supplemented by an additional
constraint which we will call the gauge condition.
The advances of the gauged DFT are the appearance of non-trivial Yang-Mills sectors.
Although T- and U-duality properties of gauged supergravities are
studied in various contexts (see for example \cite{Trigiante:2016mnt,
Samtleben:2008pe, Hassler:2023nht}), 
the doubled structure of the gauged theories are still unclear.

The purpose of this paper is to establish doubled structures of
algebroids endowed with the twisted C-bracket in the gauged DFT.
We will find that the algebroids in the gauged DFT admit Drinfel'd
double like structures. We will show that the consistency of the doubled structure leads to the physical
conditions including the strong constraint and the gauge condition in the gauged DFT.

The organization of this paper is as follows.
In Section \ref{sec:gauged_DFT}, we introduce the gauged DFT and its
gauge symmetries. The gauge algebra is governed by the twisted
C-bracket and the physical conditions in the gauged DFT are presented.
In Section \ref{sec:algebroids}, we discuss 
a geometrical implementation of the twisted C-bracket.
We will discuss a triple foliated structures in 
$(2D+n)$-dimensional product manifold $\mathcal{M}_{2D+n}$.
We will then exhibit that a metric algebroid defined by the twisted
C-bracket is constructed by three (twisted) Lie algebroids.
We find that the consistency conditions for the Courant algebroids result in
the physical conditions in the gauged DFT.
In Section \ref{sec:heterotic_gg}, we discuss 
the relations to the generalized geometry for heterotic string theories.
Section \ref{sec:conclusion} is devoted to conclusion and discussions.
A mathematical glossary is found in Appendix.

\section{Gauged double field theory and twisted C-bracket}
\label{sec:gauged_DFT}
In this section, we briefly introduce the gauged DFT and its gauge
symmetries.
We also introduce the notions of the non-Abelian gauge symmetry in DFT, the twisted
C-bracket, the strong constraint and the additional gauge constraint.
They will be key ingredients for algebroid structures discussed in Section \ref{sec:algebroids}.

We first start from the action of $O(D,D+n)$ DFT;
\begin{align}
S_0 =& \ 
\int \! \dop^{2D+n} \mathbb{X} \, e^{-2d} 
\Bigg(
\frac{1}{8} \mathcal{H}^{MN} \del_M \mathcal{H}^{KL} \del_N
 \mathcal{H}_{KL}
- \frac{1}{2} \mathcal{H}^{MN} \del_N \mathcal{H}^{KL} \del_L
 \mathcal{H}_{MK} 
\notag \\
& \qquad \qquad \qquad \qquad 
- 2 \del_M d \del_N \mathcal{H}^{MN} + 4 \mathcal{H}^{MN} \del_M d
 \del_N d
\Bigg),
\label{eq:DFT_action}
\end{align}
where $\mathcal{H}_{MN} (\mathbb{X}), \, (M,N = 1, \ldots, 2D+n)$ and $d (\mathbb{X})$ are the generalized metric and
the generalized dilaton defined in the $(2D+n)$-dimensional doubled space $\mathcal{M}_{2D+n}$ for which the
coordinate $\mathbb{X}^M$ may be decomposed into 
$\mathbb{X}^M = (\tilde{x}_{\mu}, x^{\mu}, \bar{x}_{\alpha})
, \, (\mu, \nu = 1, \ldots, D, \alpha = 1,
\ldots, n)$.
The standard parametrizations of the generalized metric and the dilaton are given by 
\begin{align}
\mathcal{H}_{MN} =& \ 
\left(
\begin{array}{ccc}
g^{\mu \nu} & - g^{\mu \rho} c_{\rho \nu} & - g^{\mu \rho} A_{\rho}{}^{\beta} \\
- g^{\nu \rho} c_{\rho \mu} & g_{\mu \nu} + c_{\rho \mu} g^{\rho
 \sigma} c_{\sigma \nu} + \kappa^{\alpha \beta} A_{\mu \alpha} A_{\nu \beta}
 & c_{\rho \mu} g^{\rho \sigma} A_{\sigma}{}^{\beta} + A_{\mu}{}^{\beta} \\
- g^{\nu \rho} A_{\rho}{}^{\alpha} & c_{\rho \nu} g^{\rho \sigma} A_{\sigma
 }{}^{\alpha} + A_{\nu}{}^{\alpha} & \kappa^{\alpha \beta} + A_{\rho}{}^{\alpha}
 g^{\rho \sigma} A_{\sigma}{}^{\beta}
\end{array}
\right),
\notag \\
e^{-2d} =& \ \sqrt{-g} e^{-2\phi},
\end{align}
where $g_{\mu \nu} (\mathbb{X}), g^{\mu \nu} (\mathbb{X})$ are a symmetric $D \times D$ matrix and
its inverse. 
An $n \times n$ symmetric constant matrix $\kappa_{\alpha \beta}$ and its
inverse $\kappa^{\alpha \beta}$ and a $D \times n$ matrix $A_{\mu
\alpha} (\mathbb{X})$ and a scalar quantity $\phi (\mathbb{X})$ have been introduced.
We have also defined the quantity
\begin{align}
c_{\mu \nu} = B_{\mu \nu} + \frac{1}{2} \kappa^{\alpha \beta} A_{\mu \alpha} A_{\nu \beta},
\end{align} 
where $B_{\mu \nu}$ is a $D \times D$ anti-symmetric matrix.
The indices $M,N, \ldots = 1, \ldots, 2D+n$ are raised and lowered by
the $O(D,D+n)$ invariant metric;
\begin{align}
\eta_{MN} = 
\left(
\begin{array}{ccc}
0 & \delta^{\mu} {}_{\nu} & 0 \\
\delta_{\mu} {}^{\nu} & 0 & 0 \\
0 & 0 & \kappa^{\alpha \beta}
\end{array}
\right),
\label{eq:invariant_metric}
\end{align}
and its inverse $\eta^{MN}$.
Note that the generalized metric $\mathcal{H}_{MN}$ is an element of
$O(D,D+n)$.
The action \eqref{eq:DFT_action} is manifestly invariant under the
$O(D,D+n)$ transformation;
\begin{align}
\mathcal{H}^{\prime MN} (\mathbb{X}') = O^M {}_P O^N {}_Q \mathcal{H}^{PQ} (\mathbb{X}),
\hspace{0.8em}
d'(\mathbb{X}') = d(\mathbb{X}),
\hspace{0.8em}
\mathbb{X}^{\prime M} = O^M {}_N \mathbb{X}^N,
\hspace{0.8em}
O \in  O (D,D+n).
\end{align}

Now we gauge a subgroup $G$ of $O(D,D+n)$ and break it down to $O(D,D)
\times G$.
This is done by introducing a constant (2,1)-tensor 
flux $F^M {}_{NK}$ such as 
\begin{align}
F^M {}_{NK} = 
\left\{
\begin{array}{ll}
F_{\alpha} {}^{\beta \gamma} & \text{if } (M,N,K) = (\alpha, \beta,
 \gamma) \\
0 & \text{else}
\end{array}
\right.
.
\label{eq:flux_constant}
\end{align}
Here $F_{\alpha} {}^{\beta \gamma}$ is the structure constant for the
gauge group $G$ whose dimension is $\dim G = n$.
The constant $F^M {}_{NK}$ must satisfy the following relations;
\begin{align}
F^{(M} {}_{PK} \eta^{N)K} = 0,
\qquad
F_{MNK} = F_{[MNK]},
\qquad
F^M {}_{N[K} F^N {}_{LP]} = 0.
\end{align}
In order to keep the gauge invariance, 
the action \eqref{eq:DFT_action} is deformed such as \cite{Hohm:2011ex}
\begin{align}
S = S_0 + \delta S,
\label{eq:gauged_action}
\end{align}
where
\begin{align}
\delta S =& \ \int \! \dop^{2D+n} \mathbb{X} \, e^{-2d} 
\Bigg(
- \frac{1}{2} 
F^{M} {}_{NK} \mathcal{H}^{NP} \mathcal{H}^{KQ} \del_P
 \mathcal{H}_{QM}
- \frac{1}{12} 
F^{M} {}_{KP} 
F^{N} {}_{LQ} \mathcal{H}_{MN} \mathcal{H}^{KL}
 \mathcal{H}^{PQ}
\notag \\
& \qquad \qquad \qquad \qquad 
- \frac{1}{4} F^{M} {}_{NK} F^N {}_{ML} \mathcal{H}^{KL}
- \frac{1}{6} F^{MNK} F_{MNK}
\Bigg).
\end{align}
The action \eqref{eq:gauged_action} is invariant under the following
gauge transformation;
\begin{align}
\delta_{\Xi} \mathcal{H}^{MN} =& \ \Xi^P \del_P \mathcal{H}^{MN} + 
(\del^M \Xi_P - \del_P \Xi^M) \mathcal{H}^{PN}
+
(\del^N \Xi_P - \del_P \Xi^N) \mathcal{H}^{MP}
- 2 \Xi^P F^{(M} {}_{PK} \mathcal{H}^{N)K},
\notag \\
\del_{\Xi} d =& \ \Xi^M \del_M d - \frac{1}{2} \del_M \Xi^M,
\label{eq:gauge_transf}
\end{align}
provided that the following physical conditions are satisfied;
\begin{align}
\eta^{MN} \del_M \del_N * = 0,
\qquad
\eta^{MN} \del_M * \del_N * = 0,
\qquad 
F^M {}_{NK} \del_M * = 0.
\label{eq:physical_conditions}
\end{align}
Here $*$ are all the quantities in DFT including the generalized metric,
the generalized dilaton and the gauge parameters $\Xi^M$.
The first condition above is just the level matching condition of closed
strings and the second is known as the strong constraint in the
context of the ordinary $O(D,D)$ DFT.
We call these the physical conditions.
The last one is specific to the gauged version of DFT
and we call it the gauge condition in the following.

The gauge transformation \eqref{eq:gauge_transf} is closed under the
 conditions \eqref{eq:physical_conditions}, namely, for an
arbitrary $O(D,D+n)$ vector $V^M$, we have
\begin{align}
[\delta_{\Xi_1}, \delta_{\Xi_2}] V^M = \delta_{[\Xi_1, \Xi_2]_F} V^M,
\end{align}
where the left-hand side is the commutator of $\delta_{\Xi_1}$ and
$\delta_{\Xi_2}$. 
In the right-hand side, we have defined the twisted C-bracket;
\begin{align}
([\Xi_1, \Xi_2]_F)^M 
= 
\Xi_1^K \del_K \Xi_2^M - \Xi_2^K \del_K \Xi_1^M
- \frac{1}{2} \eta^{MN} \eta_{KL} (\Xi_1^K \del_N \Xi_2^L - \Xi_2^K
 \del_N \Xi_1^L)
+
\Xi_2^N \Xi_1^K F^M {}_{NK}.
\label{eq:twisted_C-bracket}
\end{align}

Note that the conditions \eqref{eq:physical_conditions} are trivially
solved by quantities that depend only on $x^{\mu}$.
In this case, the action \eqref{eq:gauged_action} reduces to that of a gauged
supergravity in $D$ dimensions.
Among other things, when $D=10$ and $n=496$ and $G$ is $SO(32)$ or $E_8
\times E_8$, the theory reduces to the heterotic supergravities in ten
dimensions.

\section{Extended doubled structure for algebroids}
\label{sec:algebroids}
In this section, we introduce a geometric implementation of (twisted) Lie
algebroids in the extended doubled space $\mathcal{M}_{2D+n}$
and discuss algebras associated with it.
We show that the twisted C-bracket \eqref{eq:twisted_C-bracket} in
the gauged DFT is rewritten by the geometric quantities.
We then discuss an extended doubled structure of the metric algebroid
defined by the twisted C-bracket.
We in particular focus on the compatibility conditions of algebroids and
analyze the relation with the physical conditions in the gauged DFT.

\subsection{Lie algebroids in $\mathcal{M}_{2D+n}$}
\begin{figure}[t]
\begin{center}
\includegraphics[scale=0.5]{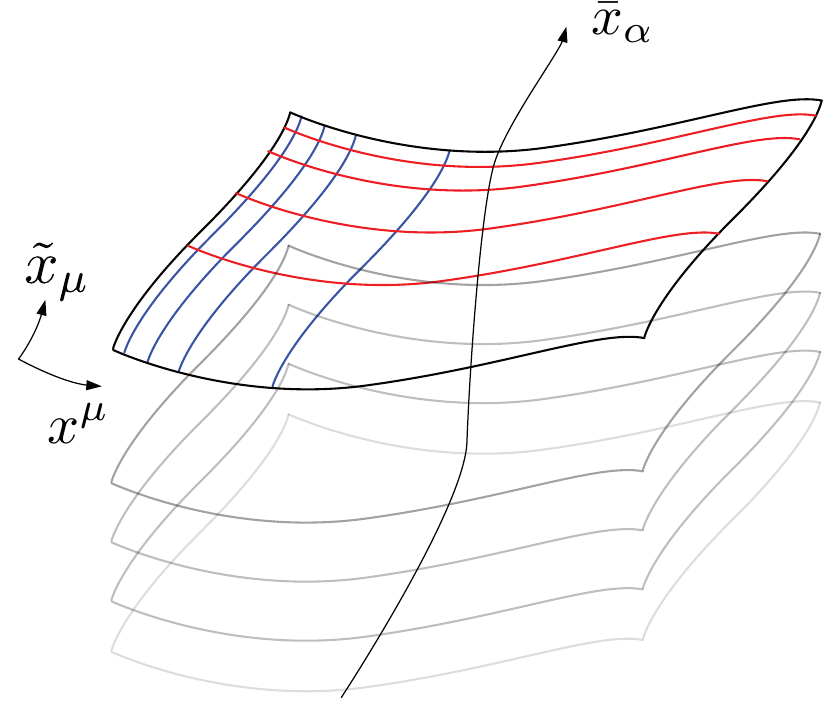}
\end{center}
\caption{
A schematic picture of the tri-foliation.
Each sheet represents $\mathcal{M}_{2D}$ specified by $\bar{x}_{\alpha} = \text{const}$.
The red and blue lines indicate spaces given by $\tilde{x}_{\mu} = \text{const.}$ and
 $x^{\mu} = \text{const}$. 
}
\label{fig:tri-foliations}
\end{figure}

We assume that the $(2D+n)$-dimensional doubled space
$\mathcal{M}_{2D+n}$ endowed with the pseudo-Riemannian metric $\eta_{MN}$ in \eqref{eq:invariant_metric} 
admits an integrable product structure.
Namely, there exists an endomorphism $\mathcal{P} : T \mathcal{M}_{2D+n}
\to T \mathcal{M}_{2D+n}$ such that $\mathcal{P}^2 = \mathbf{1}_{2D+n}$
and its associated Nijenhuis tensor vanishes.
The product structure $\mathcal{P}$ defines a rank $n$ distribution
$\bar{L}$ in $T \mathcal{M}_{2D+n}$ and its complementary pair
$\mathcal{D}$.
They corresponds to $\mathcal{P} = +1$ and $\mathcal{P} =
-1$ eigenbundles.
Then we have the decomposition of the tangent bundle $T
\mathcal{M}_{2D+n} = \mathcal{D} \oplus \bar{L}$.
Since the distributions are integrable, there exists a natural
coordinate system of 
$\mathcal{M}_{2D+n}$ ; $\mathbb{X}^M = (X^{\hat{M}}, \bar{x}_{\alpha})$
$\hat{M} = 1, \ldots, 2D$, $\alpha = 1, \ldots, n$ such
that $\mathcal{D}, \bar{L}$ are spanned by the basis $\left\{
\frac{\del}{\del X^{\hat{M}}} \right\}$ and
 $\left\{\frac{\del}{\del
\bar{x}_{\alpha}} \right\}$ 
at each point in $\mathcal{M}_{2D+n}$.
Then the structure group $O(D,D+n)$ is naturally introduced.
In particular, the integral manifolds of $\mathcal{D}$, $\bar{L}$ are defined by
slices 
$\bar{x}_{\alpha} = \text{const}.$ and 
$
X^{\hat{M}} = \text{const}$, respectively.
We further assume that the $2D$-dimensional manifold defined by
$
\bar{x}_{\alpha}
 = \text{const}.
$ is a para-hermitian manifold
admitting Born structures of DFT \cite{Freidel:2017yuv, Svoboda:2018rci,
Freidel:2018tkj}
\footnote{
From a physical point of view, 
this $2D$-dimensional manifold can also be an $L$-($\tilde{L}$-)para-hermitian manifold 
since we identify a $D$-dimensional smooth subspace determined by a
half-integrable para-hermitian structure with a physical spacetime.
}
. Due to the integrable para-hermitian structure, the
tangent space of the $2D$-dimensional manifold $\mathcal{M}_{2D}$ is
decomposed into the ones spanned by $\left\{ \frac{\del}{\del x^{\mu}} \right\}$ and
$\left\{ \frac{\del}{\del \tilde{x}_{\mu}} \right\}$ $(\mu = 1, \ldots, D)$ at each
point where 
$X^{\hat{M}} = (x^{\mu}, \tilde{x}_{\mu})$ is the natural coordinate
system in $\mathcal{M}_{2D}$.
We therefore have the decomposition
\begin{align}
T \mathcal{M}_{2D+n} = \mathcal{D} \oplus \bar{L} = L \oplus \tilde{L} \oplus \bar{L},
\end{align}
where $L, \tilde{L}$ are integrable distributions defined by the
para-hermitian structure such that $L = T\mathcal{F}$, $\tilde{L} = T
\tilde{\mathcal{F}}$.
Here the base space $\mathcal{F}$, which may be identified with the physical spacetime,
 is defined by $\tilde{x}_{\mu} =
\text{const.}$ and parameterized by $x^{\mu}$.
The other $\tilde{\mathcal{F}}$ is defined by $x^{\mu} = \text{const.}$
and parameterized by $\tilde{x}_{\mu}$.
These structures allow us to find the foliation property for $\mathcal{M}_{2D+n}$, namely, 
the coordinate of the base space is decomposed as 
$
\mathbb{X}^M = 
(\tilde{x}_{\mu}, x^{\mu}, \bar{x}_{\alpha})
$.
A hypersurface $\mathcal{M}_{2D}$ defined by $\bar{x}_{\alpha} =
\text{const}.$ contains leaves spanned by $x_{\mu},\tilde{x}^{\mu}$.
The $n$-dimensional gauge space is defined by a leaf $x^{\mu} = \text{const},
\tilde{x}_{\mu} = \text{const}.$ for which it is endowed with a metric $\kappa_{\alpha \beta}$.
These structures define a tri-foliated space (Fig \ref{fig:tri-foliations}).
Then a vector field $\Xi$ on $\mathcal{M}_{2D+n}$ is decomposed as $\Xi =
\Xi^M \del_M = X^{\mu} \del_{\mu} + \xi_{\mu} \tilde{\del}^{\mu} + a_{\alpha} \bar{\del}^{\alpha}$.

%%%%%%%
%%%%%%%

Given these properties, we define the Lie brackets in each subbundle by
\begin{align}
[X_{1}, X_{2}]_L =& \ (X_{1}^{\nu} \del_{\nu} X_{2}^{\mu} - X_{2}^{\nu} \del_{\nu} X_{1}^{\mu})
 \del_{\mu}, 
&& 
X_{1}, X_{2} \in \Gamma (L),
\notag \\
[\xi_{1}, \xi_{2}]_{\tilde{L}} =& \ (\xi_{1\nu} \tilde{\del}^{\nu} \xi_{2\mu}
 - \xi_{2\nu} \tilde{\del}^{\nu} \xi_{1\mu}) \tilde{\del}^{\mu},
&&
\xi_{1}, \xi_{2} \in \Gamma (\tilde{L}),
\notag \\
[a_1, a_2]_{\bar{L}} =& \ (a_{1 \beta}
 \bar{\del}^{\beta} a_{2 \alpha} - a_{2 \beta}
 \bar{\del}^{\beta} a_{1 \alpha}) \bar{\del}^{\alpha},
&&
a_{1}, a_{2} \in \Gamma (\bar{L}).
\label{eq:Lie_brackets}
\end{align}
For later convenience, we define 
differential operators $\dop, \tilde{\dop}, \bar{\dop}$
for a function $f$ in $\mathcal{M}_{2D+n}$;
\begin{align}
\dop f = \tilde{\del}^{\mu} f \del_{\mu} \in \Gamma (L), 
\qquad 
\tilde{\dop} f = \del_{\mu} f \tilde{\del}^{\mu} \in \Gamma (\tilde{L}), 
\qquad \bar{\dop} f = \kappa_{\alpha \beta} \bar{\del}^{\alpha} f
 \bar{\del}^{\beta} \in \Gamma (\bar{L}).
\label{eq:exterior_derivatives}
\end{align}
We now introduce totally anti-symmetric products of vector fields in $L,
\tilde{L}$ and $\bar{L}$;
\begin{align}
X^{(p)} =& \ \frac{1}{p!} X^{\mu_1 \cdots \mu_p} (\mathbb{X}) \del_{\mu_1}
 \wedge \ldots \wedge \del_{\mu_p} \in \Gamma(L^{\wedge p}),
\notag \\
\xi^{(p)} =& \ \frac{1}{p!} \xi_{\mu_1 \cdots \mu_p} (\mathbb{X})
 \tilde{\del}^{\mu_1} \wedge \cdots \wedge \tilde{\del}^{\mu_p} \in
 \Gamma (\tilde{L}^{\wedge p}),
\notag \\
a^{(p)} =& \ \frac{1}{p!} a_{\alpha_1 \cdots \alpha_p} (\mathbb{X})
 \bar{\del}^{\alpha_1} \wedge \cdots \wedge \bar{\del}^{\alpha_p} \in
\Gamma (\bar{L}^{\wedge p}),
\end{align}
where $1 \le p \le D$ for $L, \tilde{L}$ and $1 \le p \le n$ for $\bar{L}$.
The operators $\dop, \tilde{\dop}, \bar{\dop}$ act on a $p$-vector field
$\Xi^{(p)} = 
\frac{1}{p!}
\Xi^{M_1 \cdots M_p} \del_{M_1} \wedge \cdots  \wedge
\del_{M_p}$ in $T\mathcal{M}_{2D+n}$ as
\begin{align}
\dop \Xi^{(p)} =& \ \frac{1}{p!} \del_{\mu} \Xi^{M_1 \cdots M_p} (\mathbb{X}) \tilde{\del}^{\mu} \wedge
 \del_{M_1} \wedge \cdots \wedge \del_{M_p},
\notag \\
\tilde{\dop} \Xi^{(p)} =& \ \frac{1}{p!} \tilde{\del}^{\mu} \Xi^{M_1 \cdots M_p}
 (\mathbb{X}) \del_{\mu} \wedge
 \del_{M_1} \wedge \cdots \wedge \del_{M_p},
\notag \\
\bar{\dop} \Xi^{(p)} =& \ \frac{1}{p!} \kappa_{\alpha \beta} \bar{\del}^{\beta} \Xi^{M_1
 \cdots M_p} (\mathbb{X}) \bar{\del}^{\alpha} \wedge \del_{M_1} \wedge \cdots \wedge \del_{M_p}.
\end{align}
These are analogous to exterior derivatives to $p$-forms and
defined by maps that increase the rank in each wedge product.
For example, when 
$\Xi^{(p)} = 
\frac{1}{p!}
X^{(p)} = X^{\mu_1 \cdots \mu_p} \del_{\mu_1} \wedge \cdots
\wedge \del_{\mu_p} \in \Gamma (L^{\wedge p})$, we have
\begin{align}
&
\dop X^{(p)} =\ \frac{1}{p!} \del_{\mu} X^{\mu_1 \cdots \mu_p} \tilde{\del}^{\mu} \wedge
 \del_{\mu_1} \wedge \cdots \wedge \del_{\mu_p},
\notag \\
&
\tilde{\dop} X^{(p)} =\ \frac{1}{p!} \tilde{\del}^{\mu} X^{\mu_1 \cdots \mu_p} \del_{\mu}
 \wedge \del_{\mu_1} \wedge \cdots \wedge \del_{\mu_p},
\notag \\
&
\bar{\dop} X^{(p)} =\ \frac{1}{p!} \kappa_{\alpha \beta} \bar{\del}^{\beta} X^{\mu_1 \cdots
 \mu_p} \bar{\del}^{\alpha} \wedge \del_{\mu_1} \wedge \cdots \wedge \del_{\mu_p}.
\end{align}
These operators $\dop, \tilde{\dop}, \bar{\dop}$
satisfy the following properties;
\begin{align}
& \dop^2 = \tilde{\dop}^2 = \bar{\dop}^2 = 0,
\notag \\
&
\dop \tilde{\dop} + \tilde{\dop} \dop = \dop \bar{\dop} + \bar{\dop} \dop = \tilde{\dop} \bar{\dop} +
 \bar{\dop} \tilde{\dop} = 0.
\end{align}
Moreover, for a $p$-vector field $\Xi^{(p)}$ and a $q$-vector field $\Sigma^{(q)}$, we have
\begin{align}
\dop (\Xi^{(p)} \wedge \Sigma^{(q)}) = \dop \Xi^{(p)} \wedge \Sigma^{(q)} + (-)^p \Xi^{(p)} \wedge \dop \Sigma^{(q)},
\end{align}
and so on.
We next define ``inner products'' by
\begin{align}
\langle X, \xi \rangle =& \  X^{\mu} \xi_{\mu} = X (\xi) = \xi (X),
 \quad X \in \Gamma (L), \ \xi \in \Gamma (\tilde{L}),
\notag \\
\langle a_1, a_2 \rangle =& \ \kappa_{\alpha \beta}
 a^{\alpha}_1 a^{\beta}_2 = a_1 (a_2) =
 a_2 (a_1), \quad a_1, a_2 \in \Gamma (\bar{L}).
\end{align}
The other combinations vanish.
Given these inner products, $L$ and $\tilde{L}$ are dual vector bundle to each other and 
$\bar{L}$ is self-dual.
These notions enable us to introduce interior products $\tilde{\iop}, \iop, \bar{\iop}$ 
as anti-derivatives of $\tilde{\dop}, \dop, \bar{\dop}$. Namely we define 
\begin{align}
\tilde{\iop}_{\tilde{\del}^{\mu}} \del_{\nu} = \delta^{\mu} {}_{\nu},
\qquad
\iop_{\del_{\mu}} \tilde{\del}^{\nu} = \delta_{\mu} {}^{\nu},
\qquad
\bar{\iop}_{\bar{\del}^{\alpha}} \bar{\del}^{\beta} = \kappa^{\alpha
 \beta}.
\label{eq:interior_derivatives}
\end{align}
The other combinations are all zero.
They are maps that decrease the rank in each wedge products.
They act, for example, like
\begin{align}
&
\tilde{\iop}_{\xi} X^{(p)} = \frac{1}{(p-1)!}\xi_{\mu} X^{\mu \mu_2 \cdots \mu_p}
 \del_{\mu_2} \wedge \cdots 
\wedge
\del_{\mu_p},
\notag \\
&
\iop_X \xi^{(p)} = \frac{1}{(p-1)!} X^{\mu} \xi_{\mu \mu_2 \cdots \mu_p} \tilde{\del}^{\mu_2}
 \wedge \cdots \wedge \tilde{\del}^{\mu_p},
\notag \\
&
\bar{\iop}_{a_1} a_2^{(p)} = \frac{1}{(p-1)!} 
\kappa^{\alpha \alpha_1} a_{1 \alpha} a_{2 \alpha_1 \alpha_2 \cdots
 \alpha_p}
 \bar{\del}^{\alpha_2} \wedge \cdots \wedge \bar{\del}^{\alpha_p}.
\end{align}
Note that all the interior products anti-commute with all the exterior derivatives.
Then, we have the following properties;
\begin{align}
&
\tilde{\iop}^2 = \iop^2 = \bar{\iop}^2 = 0,
\notag \\
&
\tilde{\iop} \, \iop + \iop \, \tilde{\iop} = \iop \, \bar{\iop} + \bar{\iop} \, \iop = \bar{\iop} \, \tilde{\iop} +
 \tilde{\iop} \, \bar{\iop} = 0.
\end{align}
These are economically written as 
\begin{align}
\mathbf{i}_{\del_M} \del_N = \eta_{MN} = 
\left(
\begin{array}{ccc}
0 & 1 & 0 \\
1 & 0 & 0 \\
0 & 0 & \kappa^{\alpha \beta}
\end{array}
\right),
\end{align}
where $\mathbf{i} = (\tilde{\iop}, \iop, \bar{\iop})$.
For, $\Xi_1 = \Xi_1^M \del_M$, $\Xi_2^{(p)} = \frac{1}{p!} \Xi_2^{M_1 \cdots
M_p} \del_{M_1} \wedge \cdots \wedge \del_{M_p}$, we find the natural relation;
\begin{align}
\mathbf{i}_{\Xi_1} \Xi_2^{(p)} =& \
\frac{1}{p!} \sum_{s=1}^p \eta_{M_s N} \Xi_1^{N} \Xi_2^{M_1 \cdots M_s \cdots M_p}
 (-)^{s-1}
\del_{M_1} \wedge \cdots \wedge \check{\del}_{M_s} \wedge \cdots \wedge \del_{M_p}.
\end{align}
Here the symbol $\check{ \quad }$ stands for removing the corresponding part.

By using these operations, we define Lie derivatives as 
\begin{align}
&
\mathcal{L}_X = \iop_X \dop + \dop \iop_X,
\qquad
\tilde{\mathcal{L}}_{\xi} = \tilde{\iop}_{\xi} \tilde{\dop} + \tilde{\dop}
\tilde{\iop}_{\xi},
\qquad
\bar{\mathcal{L}}_{a} = \bar{\iop}_{a} \bar{\dop} + \bar{\dop} \bar{\iop}_{a},
\notag \\
&
X \in \Gamma (L),
\quad 
\xi \in \Gamma (\tilde{L}),
\quad
a \in \Gamma (\bar{L}).
\label{eq:Lie_derivatives}
\end{align}
Note that these expressions are valid only for the vector spaces that
are different from the defining space of the operators.
For example, when $\mathcal{L}_X$ defined by $X \in \Gamma (L)$ acts on
$\xi \in \Gamma (\tilde{L})$, the expression \eqref{eq:Lie_derivatives}
is valid. However, when the defining and the acting spaces are the same, 
we have the following expressions;
\begin{align}
\mathcal{L}_{X_{1}} X_{2} = [X_{1}, X_{2} ]_{L}, 
\qquad
\tilde{\mathcal{L}}_{\xi_{1}} \xi_{2} = [\xi_{1}, \xi_{2}]_{\tilde{L}},
\end{align}
where $[\cdot, \cdot]_L, [\cdot,\cdot]_{\tilde{L}}
$ are Lie brackets in $L, \tilde{L},
$ respectively.
We note that this rule does not hold in $\bar{L}$, namely, 
we have $\bar{\mathcal{L}}_{a_1} a_2 \not= [a_1, a_2]_{\bar{L}}$.
We also define ``Lie-like derivatives'' as 
\begin{align}
\bar{\mathfrak{L}}_X = \iop_X \bar{\dop} + \bar{\dop} \iop_X,
\qquad
\bar{\mathfrak{L}}_{\xi} = \tilde{\iop}_{\xi} \bar{\dop} + \bar{\dop}
 \tilde{\iop}_{\xi},
\qquad
\mathfrak{L}_{a} = \bar{\iop}_{a} \dop + \dop
 \bar{\iop}_{a},
\qquad
\tilde{\mathfrak{L}}_a = \bar{\iop}_a \tilde{\dop} +
 \tilde{\dop} \bar{\iop}_a.
\label{eq:Lie-like_derivatives}
\end{align}
They are valid in any cases discussed above.

\paragraph{Lie algebroids on $L$, $\tilde{L}$}
We next introduce Lie algebroid structures in each subbundle.
A Lie algebroid is a triple $(E, \rho, [\cdot, \cdot])$ where
$E \to M$ is a vector bundle over a base manifold $M$, $\rho :
E \to TM$ is an anchor map, and $[\cdot, \cdot]$ is the Lie bracket on $E$
satisfying the Jacobi identity 
\begin{align}
[V_1, [V_2,V_3]] + [V_2, [V_3, V_1]] + [V_3, [V_1, V_2]] = 0, \qquad V_i \ (i=1,2,3) \in \Gamma (E)
\end{align}
and a compatibility condition,
\begin{align}
[V_1, f V_2] = (\rho (V_1) \cdot f) V_2
+ f [V_1,V_2], \quad V_i \ (i=1,2) \in
 \Gamma (E).
\label{eq:bracket_anchor}
\end{align}
Here $f : M \to \mathbb{R}$ is a function on $M$.

We now discuss Lie algebroids in $L$ and $\tilde{L}$.
On the subbundle $L$, we have the Lie bracket 
$[X_{1}, X_{2}]_L = (X_{1}^{\nu} \del_{\nu} X_{2}^{\mu} - X_{2}^{\nu} \del_{\nu} X_{1}^{\mu})
 \del_{\mu}$ satisfying the Jacobi identity.
The same is true for $\tilde{L}$.
Using the 
operators $\dop, \tilde{\dop}$
in \eqref{eq:exterior_derivatives}, 
 the anchor maps
\begin{align}
\rho : L \to T \mathcal{M}_{2D+n}, 
\qquad
\tilde{\rho} : \tilde{L} \to T \mathcal{M}_{2D+n}, 
\end{align}
are generically defined through the following 
relations
;
\begin{align}
&
\dop f (X) = \rho (X) \cdot f
,
\qquad
\tilde{\dop} f (\xi) = \tilde{\rho} (\xi) \cdot f
,
\notag \\
&
X \in \Gamma (L), \quad 
\xi \in \Gamma (\tilde{L}).
\end{align}
The simplest case is that the anchor 
is given by the identity endomorphism $\rho = \mathrm{id}$.
Then $(L, \mathrm{id}, [\cdot, \cdot]_L)$ defines a Lie algebroid on a
leaf $\tilde{x}_{\mu}, \bar{x}_{\alpha} = \text{const}$.
The same is true for $(\tilde{L}, \mathrm{id},
[\cdot,\cdot]_{\tilde{L}})$ on $x^{\mu}, \bar{x}_{\alpha} = \text{const}$.

\paragraph{Twisted Lie algebroid on $\bar{L}$}
In order to discuss the algebroid on $\bar{L}$, 
we here define the notion of twisted Lie algebroids.
Given a Lie algebroid $(E,\rho, [\cdot,\cdot])$,
a twisted Lie algebroid is defined by a triple $(E, \rho,
[\cdot,\cdot]_F)$ where $[\cdot, \cdot]_F$ is the twisted bracket
defined by\footnote{The factor 2 in front of $F$ is for later convenience.}
\begin{align}
[V_1, V_2]_F = [V_1, V_2] + 2 
\iop_{V_2} \iop_{V_1} F,
\qquad
V_1, V_2 \in \Gamma (E).
\end{align}
Here 
$F \in \Gamma ( (E^* \wedge E^*) \otimes E )$
is a constant $(2,1)$-tensor and $E^*$ is the dual vector bundle of $E$.
The symbol $\iop_V$ for $V \in \Gamma (E)$ 
stands for the partial contraction using the
canonical pairing between $E$ and $E^*$.
The twisted bracket must satisfy the Jacobi identity.

We can confirm that the condition \eqref{eq:bracket_anchor} is automatically satisfied when
$(E, \rho, [\cdot, \cdot])$ is a Lie algebroid;
\begin{align}
[V_1, f V_2]_F =& \ [V_1, f V_2] + 2 
\iop_{f V_2} \iop_{V_1} F
\notag \\
=& \ \rho (V_1) 
\cdot f \,
 V_2 + f [V_1, V_2] + 2 
f \iop_{V_2} \iop_{V_1} F
\notag \\
=& \ \rho (V_1) 
\cdot f \,
 V_2 + f [V_1, V_2]_F.
\end{align}
On the other hand, the Jacobiator for the twisted bracket is calculated as 
\begin{align}
\mathrm{Jac} (V_1, V_2, V_3)_F =& \ 
  [V_1, [V_2, V_{3}]_F]_F
+ [V_2, [V_3, V_1]_F]_F
+ [V_3, [V_1, V_2]_F]_F
\notag \\
=& \ 
\mathrm{Jac} (V_1, V_2, V_3)
\notag \\
& \ 
+ 
2
\Big(
[V_1, \iop_{V_3} \iop_{V_2} F] 
+ 
\iop_{[V_2,V_3]} \iop_{V_1}
 F
+ 
2
\iop_{(\iop_{V_3} \iop_{V_2} F)} \iop_{V_1}
F
 + \text{c.p.}
\Big).
\label{eq:twisted_Jacobiator}
\end{align}
Here, $\text{Jac} (V_1, V_2, V_3)$ is the Jacobiator defined by the
ordinary Lie bracket $[\cdot, \cdot]$, which vanishes by definition, and
c.p. means the cyclic permutations.
For a twisted Lie algebroid, we demand that $\text{Jac} (V_1, V_2, V_3)_F$ vanishes at each order in $F$.
This leads to the following conditions;
\begin{align}
&
\iop_{(\iop_{V_3} \iop_{V_2} F)} \iop_{V_1}
F + \text{c.p.} = 0,
\notag \\
&
 [V_1, 
\iop_{V_3} \iop_{V_2}
 F] + 
\iop_{[V_2,V_3]} \iop_{V_1}
F + \text{c.p.} = 0.
\label{eq:condition_F_1}
\end{align}
These are constraints on $F$ for the twisted Lie algebroid.

For the $\bar{L}$ space, we have the triple $(\bar{L}, \bar{\rho},
[\cdot, \cdot]_{\bar{L}\text{-}F})$. 
Here $\bar{\rho}: \bar{L} \to T\mathcal{M}_{2D+n}$ is the anchor,
and we employ $\bar{\rho} = \mathrm{id}$ as in the case of $L, \tilde{L}$.
The twisted bracket associated with $[\cdot,\cdot]_{\bar{L}}$ is defined by
\begin{align}
[a_1, a_2]_{\bar{L}\text{-}F} = [a_1, a_2]_{\bar{L}} + 2 
\bar{\iop}_{a_2} \bar{\iop}_{a_1}
 F.
\label{eq:twisted_bracket_barL}
\end{align}
The conditions in \eqref{eq:condition_F_1} are explicitly given by
\begin{align}
&
a_{1 \alpha} a_{2 \kappa} a_{3 \tau} 
\left\{
F_{\beta} {}^{\kappa \tau} F_{\gamma} {}^{\alpha \beta} 
+
F_{\beta} {}^{\tau \alpha} F_{\gamma} {}^{\kappa \beta}
+
F_{\beta} {}^{\alpha \kappa} F_{\gamma} {}^{\tau \beta}
\right\} \bar{\del}^{\gamma} = 0,
\notag \\
&
  a_{2 \alpha} a_{3 \beta} F_{\gamma} {}^{\alpha \beta} \bar{\del}^{\gamma} a_{1 \delta} \bar{\del}^{\delta}
+ a_{3 \alpha} a_{1 \beta} F_{\gamma} {}^{\alpha \beta} \bar{\del}^{\gamma} a_{2 \delta} \bar{\del}^{\delta}
+ a_{1 \alpha} a_{2 \beta} F_{\gamma} {}^{\alpha \beta} \bar{\del}^{\gamma} a_{3 \delta} \bar{\del}^{\delta} = 0.
\end{align}
Note that the indices are raised and lowered by the metrices $\kappa_{\alpha \beta}$, $\kappa^{\alpha \beta}$.
The first line above implies that the flux $F_{\alpha} {}^{\beta \gamma}$ satisfies the Jacobi
identity. The second line means 
\begin{align}
F_{\gamma} {}^{\alpha \beta} \bar{\del}^{\gamma} * = 0
\end{align}
for any vector fields $* = a_1, a_2, a_3$.
We stress that this is nothing but the gauge condition in
\eqref{eq:physical_conditions}.

\subsection{Metric algebroid and extended double}
We next reconstruct the twisted C-bracket \eqref{eq:twisted_C-bracket}
in the geometrical languages.
Given the geometrical operations  
\eqref{eq:Lie_brackets},
\eqref{eq:exterior_derivatives}, 
\eqref{eq:interior_derivatives}, 
\eqref{eq:Lie_derivatives},
\eqref{eq:Lie-like_derivatives}, 
we find that the twisted C-bracket
\eqref{eq:twisted_C-bracket} is rewritten as 
\begin{align}
&
[\Xi_1, \Xi_2]_F 
\notag \\
&= \ 
[X_1, X_2]_L 
+ \big( \tilde{\mathcal{L}}_{\xi_1} X_2 - \tilde{\mathcal{L}}_{\xi_2} X_1 \big)
+ \big( \bar{\mathcal{L}}_{a_1} X_2 - \bar{\mathcal{L}}_{a_2} X_1 \big)
+ \frac{1}{2} \big( \tilde{\mathfrak{L}}_{a_1} a_2 - \tilde{\mathfrak{L}}_{a_2} a_1 \big)
+ \frac{1}{2} \tilde{\dop} \big( \iop_{X_1} \xi_2 - \iop_{X_2} \xi_1 \big)
\notag \\
& \
+ [\xi_1, \xi_2]_{\tilde{L}}
+ \big( \mathcal{L}_{X_1} \xi_2 - \mathcal{L}_{X_2} \xi_1 \big)
+ \big( \bar{\mathcal{L}}_{a_1} \xi_2 - \bar{\mathcal{L}}_{a_2} \xi_1 \big)
+ \frac{1}{2} \big( \mathfrak{L}_{a_1} a_2 - \mathfrak{L}_{a_2} a_1 \big)
- \frac{1}{2} \dop \big( \iop_{X_1} \xi_2 - \iop_{X_2} \xi_1 \big)
\notag \\
& \ 
+ \frac{1}{2} [a_1, a_2]_{\bar{L}}
+
\frac{1}{2}
\big( \bar{\mathcal{L}}_{a_1} a_2 - \bar{\mathcal{L}}_{a_2} a_1 \big)
+ \big( \mathcal{L}_{X_1} a_2 - \mathcal{L}_{X_2} a_1 \big)
+ \big( \tilde{\mathcal{L}}_{\xi_1} a_2 - \tilde{\mathcal{L}}_{\xi_2} a_1 \big)
\notag \\
& \ 
+ \frac{1}{2} 
\big(
\bar{\mathfrak{L}}_{X_1} \xi_2 - \bar{\mathfrak{L}}_{X_2} \xi_1
\big)
+ \frac{1}{2}
\big(
\bar{\mathfrak{L}}_{\xi_1} X_2 - \bar{\mathfrak{L}}_{\xi_2} X_1
\big)
+ \bold{i}_{\Xi_{2}} \bold{i}_{\Xi_{1}} F.
\label{eq:twisted_C-bracket_geo}
\end{align}
Here the components of vector fields are decomposed as $\Xi^M_i =
(X_i^{\mu}, \xi_{i \mu}, a_{i \alpha}), \, (i=1,2)$.
Note that $\bold{i}_{\Xi_2} \bold{i}_{\Xi_1} F  = \Xi_2^M \Xi_1^N F^P {}_{MN}
\del_P \in \Gamma (\bar{L})$ due to the condition \eqref{eq:flux_constant}.
The expression \eqref{eq:twisted_C-bracket_geo}
 is similar to the doubled structure in $O(D,D)$ DFT (see
eq.~\eqref{eq:C-bracket_DFT} in Appendix \ref{Appendix:DFT}) but here it looks rather like a ``tripled''
structure constructed out of $L, \tilde{L}$ and $\bar{L}$.
It is therefore tantalizing to clarify 
the corresponding structure of algebroids associated with
the twisted C-bracket \eqref{eq:twisted_C-bracket_geo}.

In the following, we show that the twisted C-bracket
\eqref{eq:twisted_C-bracket_geo} in its geometric form defines a metric
algebroid in $\mathcal{M}_{2D+n}$.
A metric algebroid is defined by a quadruple $(\mathcal{V}, [\cdot,\cdot]_F, \boldsymbol{\rho},
\bbra{\cdot,\cdot})$ 
where $\mathcal{V}$ is a vector bundle
over a manifold $M$
, $[\cdot,\cdot]_F$
is a skew-symmetric bracket, 
$\boldsymbol{\rho}$ is an anchor, $\bbra{\cdot,\cdot}$ is
a non-degenerate symmetric bilinear form satisfying the following
conditions;
\begin{description}
\item[Axiom M1.] $[\Xi_1, f \Xi_2]_F = f [\Xi_1, \Xi_2]_F + 
(
\boldsymbol{\rho}
(\Xi_1) \cdot f) \Xi_2 - \bbra{\Xi_1,\Xi_2} \mathcal{D} f$
\item[Axiom M2.] 
$
\boldsymbol{\rho}
(\Xi_1) \cdot  \bbra{\Xi_2,\Xi_3} = \bbra{[\Xi_1, \Xi_2]_F + \mathcal{D}
       \bbra{\Xi_1,\Xi_2}, \Xi_3} + \bbra{\Xi_2, [\Xi_1, \Xi_3]_F + \mathcal{D}  \bbra{\Xi_1,\Xi_3}}$
\end{description}
where $\Xi_1, \Xi_2, \Xi_3 \in \Gamma (\mathcal{V})$
and $\mathcal{D}$ is a map $C^{\infty} (M) \to \Gamma (\mathcal{V})$.

Now we have all the ingredients to construct the metric algebroid defined by the
twisted C-bracket.
Given Lie algebroids $(L, \rho, [\cdot,\cdot]_L)$, $(\tilde{L}, \tilde{\rho},
[\cdot, \cdot]_{\tilde{L}})$ and a twisted Lie algebroid $(\bar{L},
\bar{\rho}, [\cdot, \cdot]_{\bar{L} \text{-} F})$, 
where $L \to \mathcal{M}_{2D+n}$, $\tilde{L} \to \mathcal{M}_{2D+n}$, $\bar{L} \to \mathcal{M}_{2D+n}$ are vector bundles, 
we consider $\mathcal{V} = L \oplus \tilde{L} \oplus \bar{L}$ and an
anchor 
$
\boldsymbol{\rho}
= \rho + \tilde{\rho} + \bar{\rho} : \mathcal{V} \to
T\mathcal{M}_{2D+n}$ 
and the bracket \eqref{eq:twisted_C-bracket_geo}.
We also introduce a bilinear form 
\begin{align}
 \bbra{\Xi_1,\Xi_2} = \frac{1}{2} \bigl( \tilde{\iop}_{\xi_{1}}X_{2} +
 \iop_{X_{1}} \xi_{2}  + \bar{\iop}_{a_{1}}a_{2} \bigr).
\label{eq:bilinear_form}
\end{align}
and 
$\mathcal{D} = \dop + \tilde{\dop} + \bar{\dop}$.
In the following, we show that the quadruple 
$(\mathcal{V}, \boldsymbol{\rho}, [\cdot, \cdot]_F,\bbra{\cdot, \cdot})$ 
defined by this way indeed satisfies the axioms
of the metric algebroid.
Although the trivial example is the identity map, we leave the anchor
maps arbitrary in the following \footnote{
In general, the operator $\dop$ 
in a vector bundle and the  $\dop_0$ 
in the tangent bundle over a base manifold are distinguished.
They are related by $\dop= \rho^{*} \dop_0$ where $\rho^*$ is the adjoint of $\rho$.
}
.

\paragraph{Proof of Axiom M1}
We first study Axiom M1.
Namely, for any function $f$ in $\mathcal{M}_{2D+n}$, we show the
following condition is satisfied;
\begin{align}
	[\Xi_{1}, f\Xi_{2} ]_{F} 
	= 
	f[\Xi_{1}, \Xi_{2} ]_{F}
	+ \bigl( \boldsymbol\rho(\Xi_{1}) \cdot f \bigr)\Xi_{2}
	- \bbra{\Xi_{1}, \Xi_{2}}\mathcal{D}f.
	\label{Axiom3}
\end{align}
The $(2D+n)$-dimensional vector fields $\Xi_i \ (i=1,2)$ are given by $\Xi_{i} =
X_{i} + \xi_{i} + a_{i}\ (i=1,2)$. 
The left-hand side in \eqref{Axiom3} is decomposed as 
\begin{align}
	[\Xi_{1}, f\Xi_{2} ]_{F} 
	&= 
	[X_{1}, fX_{2} ]_{F} + [X_{1}, f\xi_{2} ]_{F} + [X_{1}, fa_{2} ]_{F} \notag\\
	&\quad
	+ [\xi_{1}, fX_{2} ]_{F} + [\xi_{1}, f\xi_{2} ]_{F} + [\xi_{1}, fa_{2} ]_{F} \notag\\
	&\quad
	+ [a_{1}, fX_{2} ]_{F} + [a_{1}, f\xi_{2} ]_{F} + [a_{1}, fa_{2} ]_{F}.
	\label{eq:all_sum_Axiom3}
\end{align}
Since $[X_{1}, X_{2} ]_{F} = [X_{1}, X_{2} ]_{L}$, 
$[\xi_{1},\xi_{2} ]_{F} = [\xi_{1}, \xi_{2} ]_{\tilde{L}}$, 
by the definition of the Lie algebroids $(L, \rho, [\cdot, \cdot]_L)$,
$(\tilde{L}, \tilde{\rho}, [\cdot,\cdot]_{\tilde{L}})$, we have 
\begin{align}
	&[X_{1}, fX_{2} ]_{F} 
	= f[X_{1}, X_{2} ]_{L} + \bigl( \rho(X_{1}) \cdot f \bigr) X_{2},
	\label{eq:all_sum_Axiom3_11}
	\\
	&[\xi_{1}, f\xi_{2} ]_{F} 
	= f[\xi_{1}, \xi_{2} ]_{\tilde{L}} + \bigl( \tilde{\rho}(\xi_{1}) \cdot f \bigr) \xi_{2}.
	\label{eq:all_sum_Axiom3_22}
\end{align}
The $ [a_{1}, fa_{2} ]_{F}$ part contains terms other than the Lie
bracket;
\begin{align}
	[a_{1}, fa_{2} ]_{F}
	&= 
	\frac{1}{2}
	[a_{1}, fa_{2} ]_{\bar{L} \text{-} F} 
	+ 
	\frac{1}{2} 
	\left( 
		\tilde{\mathfrak{L}}_{a_{1}} (fa_{2}) 
		-
		\tilde{\mathfrak{L}}_{fa_{2}} a_{1}
	\right)
	+
	\frac{1}{2}
	\left(
		\mathfrak{L}_{a_{1}} (fa_{2})
		-
		\mathfrak{L}_{fa_{2}} a_{1}
	\right)
	\notag\\
	&\quad
	+
	\frac{1}{2}
	\left(
		\bar{\mathcal{L}}_{a_{1}} (fa_{2})
		-
		\bar{\mathcal{L}}_{fa_{2}} a_{1}
	\right).
	\label{eq:all_gauge_part}
\end{align}
Here $[\cdot, \cdot]_{\bar{L}\text{-}F}$ is the twisted bracket
introduced in \eqref{eq:twisted_bracket_barL}.
Using the definitions \eqref{eq:Lie_derivatives} and 
\eqref{eq:Lie-like_derivatives}, we have
\begin{align}
&
\tilde{\mathfrak{L}}_{a_{1}} (fa_{2})
= f \tilde{\mathfrak{L}}_{a_{1}}a_{2},
&&
\mathfrak{L}_{a_{1}} (fa_{2}) 
= f \mathfrak{L}_{a_{1}}a_{2}, 
&&
\notag \\
&
\tilde{\mathfrak{L}}_{fa_{2}} a_{1}
= f \tilde{\mathfrak{L}}_{a_{2}}a_{1} 
+ \bar{\iop}_{a_{2}} a_{1} \tilde{\dop} f,
&&
\mathfrak{L}_{fa_{2}} a_{1}
=  f \mathfrak{L}_{a_{2}}a_{1} 
 + \bar{\iop}_{a_{2}} a_{1} \dop f,
&&
\notag \\
&
 \bar{\mathcal{L}}_{fa_{2}} a_{1}
= f \bar{\mathcal{L}}_{a_{2}} a_{1}
+ \bar{\iop}_{a_{2}} a_{1} \bar{\dop} f .
\end{align}
Note that the interior products anti-commute with the basis of vectors.
Then we have
\begin{align}
[a_{1}, fa_{2} ]_{F}
	&=
	\frac{1}{2}
	f\Bigl(
	[a_{1}, a_{2} ]_{\bar{L} \text{-} F} 
	+ 
	(
		\tilde{\mathfrak{L}}_{a_{1}}a_{2}
		- \tilde{\mathfrak{L}}_{a_{2}}a_{1} 
	)
	+
	(
		\mathfrak{L}_{a_{1}}a_{2}
		- \mathfrak{L}_{a_{2}}a_{1} 
	)
	+ 
	(
		\bar{\mathcal{L}}_{a_{1}} a_{2}
		- \bar{\mathcal{L}}_{a_{2}} a_{1}
	)
	\Bigr)
	\notag\\
	&\quad
	+ 
	\frac{1}{2} 
	\left( 
	 	\bigl( \bar{\rho}(a_{1}) \cdot f \bigr) a_{2}
		- \bar{\iop}_{a_{2}} a_{1} \tilde{\dop} f
		- \bar{\iop}_{a_{2}} a_{1} \dop f
		+ \bigl( \bar{\rho}(a_{1}) \cdot f \bigr) a_{2}
		- \bar{\iop}_{a_{2}} a_{1} \bar{\dop} f 
	\right) 
	\notag\\
	&=
	f [a_{1}, a_{2}]_{F}
	+ \bigl( \bar{\rho}(a_{1}) \cdot f \bigr) a_{2}
	- \frac{1}{2} \bar{\iop}_{a_{2}} a_{1} \mathcal{D} f,
	\label{eq:all_sum_Axiom3_33}
\end{align}
where we have used the relation 
$\bar{\iop}_{a_{1}} (\bar{\dop} f) a_{2} = \bigl( \bar{\rho}(a_{1}) \cdot f \bigr)
a_{2}$ and the fact that $(\bar{L}, \bar{\rho},
[\cdot,\cdot]_{\bar{L} \text{-} F})$ is a twisted Lie algebroid.
Next, the $[X_1, f \xi_2]_F$ term is expanded as
\begin{align}
	[X_{1}, f\xi_{2}]_{F}
	=
	- \tilde{\mathcal{L}}_{f\xi_{2}} X_{1}
	+ \frac{1}{2} \tilde{\dop} \iop_{X_{1}} (f\xi_{2})
	+ \mathcal{L}_{X_{1}} (f\xi_{2})
	- \frac{1}{2} \dop \iop_{X_{1}} (f\xi_{2})
	+\frac{1}{2}
	\left(
		\bar{\mathfrak{L}}_{X_{1}} (f\xi_{2})
		-
		\bar{\mathfrak{L}}_{f\xi_{2}} X_{1}
	\right).
	\label{eq:gauge_all_12}
\end{align}
By using the following relations
\begin{align}
\bar{\mathfrak{L}}_{X_{1}} (f\xi_{2})
= 
f \bar{\mathfrak{L}}_{X_{1}} (\xi_{2}),
\qquad
\bar{\mathfrak{L}}_{f\xi_{2}} X_{1}
=
f \bar{\mathfrak{L}}_{\xi_{2}} X_{1}
+ \tilde{\iop}_{\xi_{2}} X_{1} \bar{\dop} f,
\end{align}
we therefore obtain
\begin{align}
[X_{1}, f\xi_{2}]_{F}
=
f [X_{1}, \xi_{2}]_{F} + \bigr( \rho(X_{1}) \cdot f \bigl) \xi_{2} - \frac{1}{2} \iop_{X_{1}} \xi_{2} \mathcal{D} f.
\label{eq:all_sum_Axiom3_12}
\end{align}
Similarly, we have
\begin{align}
[\xi_{1}, fX_{2}]_{F}
&= 
f [\xi_{1}, X_{2}]_{F} + \bigr( \tilde{\rho}(\xi_{1}) \cdot f \bigl)
 X_{2} - \frac{1}{2} \tilde{\iop}_{\xi_{1}} X_{2} \mathcal{D} f,
\notag \\
[X_{1}, fa_{2}]_{F} 
&= 
f[X_{1}, a_{2}]_{F} + \bigl( \rho(X_{1}) \cdot f \bigr) a_{2},
\notag \\
[\xi_{1}, fa_{2}]_{F} 
& = 
f[\xi_{1}, a_{2}]_{F} 
+ \bigl( \tilde{\rho}(\xi_{1}) \cdot f \bigr) a_{2},
\notag \\
[a_{1}, fX_{2}]_{F} 
&= 
f[a_{1}, X_{2}]_{F} + \bigl( \bar{\rho}(a_{1}) \cdot  f \bigr) X_{2},
\notag \\
[a_{1}, f\xi_{2}]_{F}
&= f[a_{1}, \xi_{2}]_{F} + \bigl( \bar{\rho}(a_{1}) \cdot  f \bigr) \xi_{2}.
\end{align}
Collecting all together, we find
\begin{align}
[\Xi_{1}, f\Xi_{2} ]_{F} 
&=
f[\Xi_{1}, \Xi_{2} ]_{F} 
+ \bigl( \boldsymbol\rho(\Xi_{1}) \cdot f \bigr) \Xi_{2} 
- \bbra{\Xi_{1}, \Xi_{2}} \mathcal{D} f.
\end{align}
Then we have proved Axiom M1.

\paragraph{Proof of Axiom M2}
In order to study Axiom M2, it is convenient to introduce the
following quantity;
\begin{align}
	T_{F}(\Xi_{1}, \Xi_{2}, \Xi_{3})
	=& \
	\frac{1}{3}
	\bigl(
		\bbra{ [\Xi_{1}, \Xi_{2} ]_{F}, \Xi_{3}} + \rm{c.p.}
	\bigr),
	\label{def:TF}
\end{align}
where c.p. is the cyclic permutations.
Now we decompose the bracket into the terms in $L \oplus \tilde{L}$ and
$\bar{L}$ parts;
\begin{align}
	[\Xi_{1}, \Xi_{2} ]_{F}
	&= [e_{1}, e_{2}]_{\sf C}
	\notag\\
	&\quad
	+ 
	\left(
		\bar{\mathcal{L}}_{a_{1}} X_{2}
		- 
		\bar{\mathcal{L}}_{a_{2}} X_{1}
	\right) 
	+ 
	\frac{1}{2} 
	\left( 
		\tilde{\mathfrak{L}}_{a_{1}} a_{2} 
		-
		\tilde{\mathfrak{L}}_{a_{2}} a_{1}
	\right)
	\notag\\
	&\quad
	+
	\left(
		\bar{\mathcal{L}}_{a_{1}} \xi_{2}
		-
		\bar{\mathcal{L}}_{a_{2}} \xi_{1}
	\right)
	+
	\frac{1}{2}
	\left(
		\mathfrak{L}_{a_{1}} a_{2}
		-
		\mathfrak{L}_{a_{2}} a_{1}
	\right)
	\notag\\
	&\quad
	+
	\frac{1}{2}
	[a_{1}, a_{2}]_{\bar{L}}
	+
	\frac{1}{2}
	\left(
		\bar{\mathcal{L}}_{a_{1}} a_{2}
		-
		\bar{\mathcal{L}}_{a_{2}} a_{1}
	\right)
	+
	\left(
		\mathcal{L}_{X_{1}} a_{2}
		-
		\mathcal{L}_{X_{2}} a_{1}
	\right)
	+
	\left(
		\tilde{\mathcal{L}}_{\xi_{1}} a_{2}
		-
		\tilde{\mathcal{L}}_{\xi_{2}} a_{1}
	\right) 
	\notag\\
	&\quad
	+
	\frac{1}{2}
	\left(
	\bar{\mathfrak{L}}_{X_{1}} \xi_{2}
	-
	\bar{\mathfrak{L}}_{X_{2}} \xi_{1}
	\right)
	+
	\frac{1}{2}
	\left(
		\bar{\mathfrak{L}}_{\xi_{1}} X_{2}
		-
		\bar{\mathfrak{L}}_{\xi_{2}} X_{1}
	\right)
+ \bold{i}_{\Xi_{2}} \bold{i}_{\Xi_{1}} F
.
\end{align}
Here $[\cdot, \cdot]_{\sf C}$ is the C-bracket for the ungauged DFT (see
\eqref{eq:C-bracket_DFT} in Appendix \ref{Appendix:DFT}) and 
$e_{i} = X_i + \xi_i \in\Gamma(L \oplus \tilde{L}), \, (i=1,2)$.
Then we have
\begin{align}
	\bbra{ [\Xi_{1}, \Xi_{2} ]_{F}, \Xi_{3}} 
	&=
	\bbra{ [e_{1}, e_{2}]_{\sf C} , X_{3} + \xi_{3}} 
	+
	\bbra{A + B + C, X_{3} + \xi_{3} +a_{3}}
	+ 
	\frac{1}{2} \bold{i}_{\Xi_{3}} \bold{i}_{\Xi_{2}} \bold{i}_{\Xi_{1}} F
	\notag\\
	&=
	\bbra{[e_{1}, e_{2}]_{\sf C} , X_{3} + \xi_{3}} 
	+
	\frac{1}{2}
	\bigl\{
		\tilde{\iop}_{\xi_{3}} A + \iop_{X_{3}} B + \bar{\iop}_{a_{3}} C
	\bigr\}
	+
	\frac{1}{2} \bold{i}_{\Xi_{3}} \bold{i}_{\Xi_{2}} \bold{i}_{\Xi_{1}} F,
\label{eq:sum_pbra_F}
\end{align}
where we have 
introduced the notation $\bold{i}_{\Xi_3} \bold{i}_{\Xi_2} \bold{i}_{\Xi_1}
F = \Xi_3^P \Xi_2^M \Xi_1^N \eta_{PQ} F^Q {}_{MN}$.
We have also
defined the following quantities;
\begin{align}
	A 
	&=
	\left(
		\bar{\mathcal{L}}_{a_{1}} X_{2}
		- 
		\bar{\mathcal{L}}_{a_{2}} X_{1}
	\right) 
	+ 
	\frac{1}{2} 
	\left( 
		\tilde{\mathfrak{L}}_{a_{1}} a_{2} 
		-
		\tilde{\mathfrak{L}}_{a_{2}} a_{1}
	\right),
	\\
	B 
	&=
	\left(
		\bar{\mathcal{L}}_{a_{1}} \xi_{2}
		-
		\bar{\mathcal{L}}_{a_{2}} \xi_{1}
	\right)
	+
	\frac{1}{2}
	\left(
		\mathfrak{L}_{a_{1}} a_{2}
		-
		\mathfrak{L}_{a_{2}} a_{1}
	\right),
	\\
	C
	&=
	\frac{1}{2}
	[a_{1}, a_{2}]_{\bar{L}}
	+
	\frac{1}{2}
	\left(
		\bar{\mathcal{L}}_{a_{1}} a_{2}
		-
		\bar{\mathcal{L}}_{a_{2}} a_{1}
	\right)
	+
	\left(
		\mathcal{L}_{X_{1}} a_{2}
		-
		\mathcal{L}_{X_{2}} a_{1}
	\right)
	+
	\left(
		\tilde{\mathcal{L}}_{\xi_{1}} a_{2}
		-
		\tilde{\mathcal{L}}_{\xi_{2}} a_{1}
	\right)
	\notag\\
	&\quad
	+
	\frac{1}{2}
	\left(
	\bar{\mathfrak{L}}_{X_{1}} \xi_{2}
	-
	\bar{\mathfrak{L}}_{X_{2}} \xi_{1}
	\right)
	+
	\frac{1}{2}
	\left(
		\bar{\mathfrak{L}}_{\xi_{1}} X_{2}
		-
		\bar{\mathfrak{L}}_{\xi_{2}} X_{1}
	\right).
\end{align}
The first term $\bbra{[e_{1}, e_{2}]_{\sf C} , X_{3} + \xi_{3}}$ in 
\eqref{eq:sum_pbra_F} is evaluated as \cite{Mori:2019slw}
\begin{align}
&
\bbra{[e_{1}, e_{2}]_{\sf C} , X_{3} + \xi_{3}}
\notag\\
&=
T(e_{1}, e_{2}, e_{3})
\notag\\
&\quad
+ \frac{1}{4}\rho(X_{1}) \cdot \tilde{\iop}_{\xi_{3}}X_{2} 
+ \frac{1}{4}\rho(X_{1}) \cdot \tilde{\iop}_{\xi_{2}}X_{3}
+ \frac{1}{4}\tilde{\rho}(\xi_{1}) \cdot \tilde{\iop}_{\xi_{3}}X_{2} 
+ \frac{1}{4}\tilde{\rho}(\xi_{1}) \cdot \tilde{\iop}_{\xi_{2}}X_{3}
\notag\\
&\quad
-\frac{1}{4}\rho(X_{2}) \cdot \tilde{\iop}_{\xi_{1}}X_{3} 
-\frac{1}{4}\rho(X_{2}) \cdot \tilde{\iop}_{\xi_{3}}X_{1}
-\frac{1}{4}\tilde{\rho}(\xi_{2}) \cdot \tilde{\iop}_{\xi_{1}}X_{3}
-\frac{1}{4}\tilde{\rho}(\xi_{2}) \cdot \tilde{\iop}_{\xi_{3}}X_{1}.
\label{eq:ODD_pbra}
\end{align}
where we have defined
\begin{align}
	T(e_{1}, e_{2}, e_{3})
	=
	\frac{1}{3}
	\bigl(
		\bbra{ [e_{1}, e_{2} ]_{\sf C}, e_{3}} + \rm{c.p.}
	\bigr).
\end{align}
Each part in the second term in \eqref{eq:sum_pbra_F} is evaluated as
\begin{align}
	\tilde{\iop}_{\xi_{3}}A
	&=
	\frac{1}{2} \bar{\rho}(a_{1}) \cdot \tilde{\iop}_{\xi_{3}} X_{2}
	- \frac{1}{2} \bar{\rho}(a_{2}) \cdot \tilde{\iop}_{\xi_{3}} X_{1}
	+ \frac{1}{2} \tilde{\iop}_{\xi_{3}} \bar{\iop}_{a_{1}} \bar{\dop} X_{2}
	- \frac{1}{2} \tilde{\iop}_{\xi_{3}} \bar{\iop}_{a_{2}} \bar{\dop} X_{1}
	\notag\\
	&\quad
	- \frac{1}{2} \iop_{X_{2}}\bar{\iop}_{a_{1}}\bar{\dop}\xi_{3}
	+ \frac{1}{2} \iop_{X_{1}}\bar{\iop}_{a_{2}}\bar{\dop}\xi_{3}
	+ \frac{1}{2} \tilde{\iop}_{\xi_{3}} \bar{\iop}_{a_{1}} \tilde{\dop} a_{2} 
	- \frac{1}{2} \tilde{\iop}_{\xi_{3}} \bar{\iop}_{a_{2}} \tilde{\dop} a_{1},
\notag \\
	\iop_{X_{3}}B
	&=
	\frac{1}{2} \bar{\rho}(a_{1}) \cdot \iop_{X_{3}} \xi_{2}
	- \frac{1}{2} \bar{\rho}(a_{2}) \cdot \iop_{X_{3}} \xi_{1}
	+ \frac{1}{2} \iop_{X_{3}} \bar{\iop}_{a_{1}} \bar{\dop} \xi_{2}
	- \frac{1}{2} \iop_{X_{3}} \bar{\iop}_{a_{2}} \bar{\dop} \xi_{1}
	\notag\\
	&\quad
	- \frac{1}{2} \tilde{\iop}_{\xi_{2}}\bar{\iop}_{a_{1}}\bar{\dop}X_{3}
	+ \frac{1}{2} \tilde{\iop}_{\xi_{1}}\bar{\iop}_{a_{2}}\bar{\dop}X_{3}
	+ \frac{1}{2} \iop_{X_{3}} \bar{\iop}_{a_{1}} \dop a_{2} 
	- \frac{1}{2} \iop_{X_{3}} \bar{\iop}_{a_{2}} \dop a_{1},
	\notag\\
	\bar{\iop}_{a_{3}}C
	&=
	\frac{1}{2} \rho(X_{1}) \cdot \bar{\iop}_{a_{2}}a_{3}
	- \frac{1}{2} \rho(X_{2}) \cdot \bar{\iop}_{a_{1}}a_{3}
	+ \frac{1}{2} \tilde{\rho}(\xi_{1}) \cdot \bar{\iop}_{a_{2}}a_{3}
	\notag\\
	&\quad
	- \frac{1}{2} \tilde{\rho}(\xi_{2}) \cdot \bar{\iop}_{a_{1}}a_{3}
	+ \frac{1}{2} \bar{\rho}(a_{1}) \cdot \bar{\iop}_{a_{2}} a_{3}
	- \frac{1}{2} \bar{\rho}(a_{2}) \cdot \bar{\iop}_{a_{1}} a_{3}
	\notag\\
	&\quad
	+ \frac{1}{2} \bar{\iop}_{a_{3}} \iop_{X_{1}} \dop a_{2}
	- \frac{1}{2} \bar{\iop}_{a_{3}} \iop_{X_{2}} \dop a_{1}
	- \frac{1}{2} \bar{\iop}_{a_{2}} \iop_{X_{1}} \dop a_{3}
	+ \frac{1}{2} \bar{\iop}_{a_{1}} \iop_{X_{2}} \dop a_{3}
	\notag\\
	&\quad
	+ \frac{1}{2} \bar{\iop}_{a_{3}} \tilde{\iop}_{\xi_{1}} \tilde{\dop} a_{2}
	- \frac{1}{2} \bar{\iop}_{a_{2}} \tilde{\iop}_{\xi_{1}} \tilde{\dop} a_{3}
	- \frac{1}{2} \bar{\iop}_{a_{3}} \tilde{\iop}_{\xi_{2}} \tilde{\dop} a_{1}
	+ \frac{1}{2} \bar{\iop}_{a_{1}} \tilde{\iop}_{\xi_{2}} \tilde{\dop} a_{3}
	\notag\\
	&\quad
	+ \frac{1}{2}\bar{\iop}_{a_{3}} \bar{\iop}_{a_{1}} \bar{\dop} a_{2}
	- \frac{1}{2}\bar{\iop}_{a_{3}} \bar{\iop}_{a_{2}} \bar{\dop} a_{1}
	- \frac{1}{2} \bar{\iop}_{a_{2}} \bar{\iop}_{a_{1}} \bar{\dop} a_{3}
	+ \frac{1}{2}\bar{\iop}_{a_{3}} \iop_{X_{1}} \bar{\dop} \xi_{2}
	- \frac{1}{2} \bar{\iop}_{a_{3}} \iop_{X_{2}} \bar{\dop} \xi_{1}
	\notag\\
	&\quad
	+ \frac{1}{2} \bar{\iop}_{a_{3}} \tilde{\iop}_{\xi_{1}} \bar{\dop}  X_{2}
	-  \frac{1}{2} \bar{\iop}_{a_{3}} \tilde{\iop}_{\xi_{2}} \bar{\dop} X_{1}.
\end{align}
Here we have used the relations
\begin{align}
\bar{\iop}_{a_{3}} [a_{1}, a_{2}]_{\bar{L}}
&= 
\bar{\iop}_{a_{1}} \bar{\dop} \bar{\iop}_{a_{2}} a_{3}
- \bar{\iop}_{a_{2}} \bar{\dop} \bar{\iop}_{a_{1}} a_{3}
- \bar{\iop}_{a_{2}} \bar{\iop}_{a_{1}} \bar{\dop} a_{3}
,
\notag\\
&= 
\bar{\rho}(a_{1}) \cdot \bar{\iop}_{a_{2}} a_{3}
- \bar{\rho}(a_{2}) \cdot \bar{\iop}_{a_{1}} a_{3}
- \bar{\iop}_{a_{2}} \bar{\iop}_{a_{1}} \bar{\dop} a_{3},
\notag \\
\tilde{\iop}_{\xi_{3}}\bar{\iop}_{a_{1}}\bar{\dop}X_{2}
&= 
\bar{\rho}(a_{1}) \cdot \tilde{\iop}_{\xi_{3}} X_{2}
-
\iop_{X_{2}}\bar{\iop}_{a_{1}}\bar{\dop}\xi_{3}.
\label{eq:anchor_dop}
\end{align}
Note that the last one follows from the identities;
\begin{align}
\bar{\rho}(a_{1}) \cdot \tilde{\iop}_{\xi_{3}} X_{2} 
=
\bar{\iop}_{a_{1}}\bar{\dop} \tilde{\iop}_{\xi_{3}} X_{2}
=
\bar{\mathcal{L}}_{a_{1}} \tilde{\iop}_{\xi_{3}} X_{2}
=
\tilde{\iop}_{\xi_{3}}\bar{\iop}_{a_{1}}\bar{\dop}X_{2} +
\iop_{X_{2}}\bar{\iop}_{a_{1}}\bar{\dop}\xi_{3}.
\end{align}
We find that the similar relations also hold for the following quantities,
\begin{align}
	% \mathcal{L}_{X_{1}} \bar{\iop}_{a_{1}}a_{2}
	% &=
	% \iop_{X_{1}} \dop \bar{\iop}_{a_{1}}a_{2}
	% =
	% \rho(X_{1}) \cdot \bar{\iop}_{a_{1}}a_{2}
	% =
	% \bar{\iop}_{a_{2}} \iop_{X_{1}} \dop a_{1}
	% +
	% \bar{\iop}_{a_{1}} \iop_{X_{1}} \dop a_{2}, \\
	% \tilde{\mathcal{L}}_{\xi_{1}} \bar{\iop}_{a_{1}}a_{2}
	% &=
	% \tilde{\iop}_{\xi_{1}} \dop \bar{\iop}_{a_{1}}a_{2}
	% =
	% \tilde{\rho}(\xi_{1}) \cdot \bar{\iop}_{a_{1}}a_{2}
	% =
	% \bar{\iop}_{a_{2}} \tilde{\iop}_{\xi_{1}} \tilde{\dop} a_{1}
	% +
	% \bar{\iop}_{a_{1}} \tilde{\iop}_{\xi_{1}} \tilde{\dop} a_{2}.
%
\mathcal{L}_{X_{1}} \bar{\iop}_{a_{2}}a_{3}
&=
\iop_{X_{1}} \dop \bar{\iop}_{a_{2}}a_{3}
=
\rho(X_{1}) \cdot \bar{\iop}_{a_{2}}a_{3}
=
\bar{\iop}_{a_{3}} \iop_{X_{1}} \dop a_{2}
+
\bar{\iop}_{a_{2}} \iop_{X_{1}} \dop a_{3}, \\
\tilde{\mathcal{L}}_{\xi_{1}} \bar{\iop}_{a_{2}}a_{3}
&=
\tilde{\iop}_{\xi_{1}} \dop \bar{\iop}_{a_{2}}a_{3}
=
\tilde{\rho}(\xi_{1}) \cdot \bar{\iop}_{a_{2}}a_{3}
=
\bar{\iop}_{a_{3}} \tilde{\iop}_{\xi_{1}} \tilde{\dop} a_{2}
+
\bar{\iop}_{a_{2}} \tilde{\iop}_{\xi_{1}} \tilde{\dop} a_{3}.
\end{align}
Collecting all together, we find
\begin{align}
	T_{F} (\Xi_{1}, \Xi_{2}, \Xi_{3})
	&=
	T(e_{1}, e_{2}, e_{3})
	\notag\\
	&\quad
	+ \frac{1}{4} 
		\Bigl\{
		\bigl(
		\tilde{\iop}_{\xi_{2}} \bar{\iop}_{a_{3}}  \bar{\dop} X_{1}
		+ \iop_{X_{2}} \bar{\iop}_{a_{3}} \bar{\dop} \xi_{1}
		+ \bar{\iop}_{a_{2}} \iop_{X_{3}} \dop a_{1}
		+ \bar{\iop}_{a_{2}} \tilde{\iop}_{\xi_{3}} \tilde{\dop} a_{1}
		+ \bar{\iop}_{a_{2}} \bar{\iop}_{a_{3}} \bar{\dop} a_{1}
		\bigr) 
	\notag\\
	&\qquad\qquad
		- 
		\bigl(
		\tilde{\iop}_{\xi_{3}} \bar{\iop}_{a_{2}} \bar{\dop} X_{1}
		+ \iop_{X_{3}} \bar{\iop}_{a_{2}} \bar{\dop} \xi_{1}
		+ \bar{\iop}_{a_{3}} \iop_{X_{2}} \dop a_{1}
		+ \bar{\iop}_{a_{3}} \tilde{\iop}_{\xi_{2}} \tilde{\dop} a_{1}
		\bigr)
		+ \mathrm{c.p.}
		\Bigr\}
	\notag\\
	&\quad
	+
	\frac{1}{2} \bold{i}_{\Xi_{3}} \bold{i}_{\Xi_{2}} \bold{i}_{\Xi_{1}} F
	\label{eq:Lemma3.2_Fbra}
\end{align}
and therefore this results in
\begin{align}
	\bbra{ [\Xi_{1}, \Xi_{2} ]_{F}, \Xi_{3}} 
	&=
	T_{F} (\Xi_{1}, \Xi_{2}, \Xi_{3})
	+ \frac{1}{2}\boldsymbol\rho(\Xi_{1}) \cdot \bbra{ \Xi_{3}, \Xi_{2}}
	- \frac{1}{2}\boldsymbol\rho(\Xi_{2}) \cdot \bbra{\Xi_{1}, \Xi_{3}}
	+ \frac{1}{2} \bold{i}_{\Xi_{3}} \bold{i}_{\Xi_{2}} \bold{i}_{\Xi_{1}} F
	.
	\label{eq:lemma_3.2_alt}
\end{align}
By summing up the equation \eqref{eq:lemma_3.2_alt} and the ones with 2
and 3 interchanged, we have 
\begin{align}
	\boldsymbol\rho (\Xi_{1}) \cdot \bbra{\Xi_{2}, \Xi_{3}}
	&=
	\bbra{[\Xi_{1}, \Xi_{2} ]_{F}, \Xi_{3}} 
	+
	\bbra{ [\Xi_{1}, \Xi_{3} ]_{F}, \Xi_{2}}
	\notag\\
	&\quad
	+ 
	\frac{1}{2}\boldsymbol\rho(\Xi_{2}) \cdot \bbra{\Xi_{1}, \Xi_{3}}
	+
	\frac{1}{2}\boldsymbol\rho(\Xi_{3}) \cdot \bbra{\Xi_{1}, \Xi_{2}}.
	\label{eq:axiom5_tochu1}
\end{align}
Since we have $\boldsymbol\rho = \rho + \tilde{\rho} + \bar{\rho}$,
the third term in the right-hand side in \eqref{eq:axiom5_tochu1} becomes
\begin{align}
	\frac{1}{2} \boldsymbol\rho (\Xi_{2}) \cdot \bbra{\Xi_{1}, \Xi_{3}}
	&=
	\frac{1}{2} \rho (X_{2}) \cdot \bbra{\Xi_{1}, \Xi_{3}}
	+
	\frac{1}{2} \tilde{\rho} (\xi_{2}) \cdot \bbra{\Xi_{1}, \Xi_{3}}
	+
	\frac{1}{2} \bar{\rho} (a_{2}) \cdot \bbra{\Xi_{1}, \Xi_{3}}
	\notag\\
	&=
	\frac{1}{2}
	\bigl\{
		\iop_{X_{2}} \dop \bbra{\Xi_{1}, \Xi_{3}}
		+
		\tilde{\iop}_{\xi_{2}} \tilde{\dop} \bbra{\Xi_{1}, \Xi_{3}} 
		+
		\bar{\iop}_{a_{2}} \bar{\dop} \bbra{\Xi_{1}, \Xi_{3}}
	\bigr\}
	\notag\\
	&=
	\bbra{\mathcal{D} \bbra{\Xi_{1}, \Xi_{3}}, \Xi_{2}}.
\end{align}
The same holds for the fourth term.
Then finally we obtain 
\begin{align}
	\boldsymbol\rho (\Xi_{1}) \cdot \bbra{\Xi_{2}, \Xi_{3}}
	&=
	\bbra{[\Xi_{1}, \Xi_{2} ]_{F} + \mathcal{D} \bbra{\Xi_{1}, \Xi_{2}}, \Xi_{3}} 
	+
	\bbra{ [\Xi_{1}, \Xi_{3} ]_{F} + \mathcal{D} \bbra{\Xi_{1}, \Xi_{3}}, \Xi_{2}}.
\end{align}
This proves Axiom M2.

\subsection{Courant algebroid and extended doubled structure}

We next study the extended doubled structure for a Courant algebroid defined
by the twisted C-bracket \eqref{eq:twisted_C-bracket_geo}.
In the case of ungauged DFT, the C-bracket defines a Courant algebroid provided
that the underlying Lie algebroids satisfy the condition of the Lie
bialgebroid \cite{LiWeXu}.
It has been found that this condition is nothing but the strong
constraint in DFT \cite{Mori:2019slw}.

In the following, we look for conditions that the bracket
\eqref{eq:twisted_C-bracket_geo} and the Lie algebroids on $L, \tilde{L},
\bar{L}$ define a Courant algebroid.

\paragraph{Courant algebroid}
In addition to Axioms M1 and M2 in the metric algebroid, a Courant
algebroid is defined by a quadruple $(\mathcal{V}, \boldsymbol\rho,
[\cdot,\cdot]_F, \bbra{\cdot,\cdot})$ satisfying the following Axioms;
\begin{description}
\item[Axiom C1.] 
For any $\Xi_1, \Xi_2 \in \Gamma (\mathcal{V})$, we have
\begin{align}
\boldsymbol{\rho}([\Xi_{1}, \Xi_{2}]_{F}) =  [ \boldsymbol{\rho}(\Xi_{1}), \boldsymbol{\rho}(\Xi_{2})],
\end{align}
where $[\cdot,\cdot]$ in the right-hand side is the Lie bracket in 
$\mathcal{V}$.
\item[Axiom C2.]
For any functions $f,g$, we have
\begin{align}
\bbra{\mathcal{D}f , \mathcal{D}g} = 0
\label{def:Axiom4}
\end{align}
\item[Axiom C3.]
For any $\Xi_1, \Xi_2, \Xi_3 \in \Gamma (\mathcal{V})$, we have
\begin{align}
[[\Xi_{1}, \Xi_{2} ]_{F},\Xi_{3} ]_{F} + \mathrm{c.p.} = \mathcal{D} T_{F} ( \Xi_{1}, \Xi_{2}, \Xi_{3})
\end{align}
\end{description}

We examine Axioms C1-C3 for $\mathcal{V} = L \oplus \tilde{L} \oplus
\bar{L} \to \mathcal{M}_{2D+n}$, 
the anchor $\boldsymbol{\rho} = \rho + \tilde{\rho} + \bar{\rho}$,
the twisted C-bracket \eqref{eq:twisted_C-bracket_geo},
 and the bilinear form \eqref{eq:bilinear_form}.

\paragraph{Proof of Axiom C1}
Axiom C1 implies the following relation,
\begin{align}
\boldsymbol\rho([\Xi_{1}, \Xi_{2}]_{F}) \cdot f 
-
[ \boldsymbol\rho(\Xi_{1}),
 \boldsymbol\rho(\Xi_{2})] \cdot f = 0,
\label{eq:Axiom2_act}
\end{align}
where $f$ is an arbitrary function on $\mathcal{M}_{2D+n}$.
The first term in the left-hand side in \eqref{eq:Axiom2_act} is
expanded as
\begin{align}
	\boldsymbol\rho( [\Xi_{1}, \Xi_{2}]_{F} ) \cdot f
	&=
 	\rho( [X_{1}, X_{2}]_{L} ) \cdot f
	+ 
	%\left( 
		\rho( \tilde{\mathcal{L}}_{\xi_{1}} X_{2} ) \cdot f
		- 
		\rho( \tilde{\mathcal{L}}_{\xi_{2}} X_{1} ) \cdot f
	%\right)
	+ 
	%\left(
		\rho( \mathcal{L}_{a_{1}} X_{2} ) \cdot f
		- 
		\rho( \mathcal{L}_{a_{2}} X_{1} ) \cdot f
	%\right) 
	\notag\\
	&\qquad
	+ 
	\frac{1}{2} 
	\left( 
		\rho( \tilde{\mathfrak{L}}_{a_{1}} a_{2}  ) \cdot f
		-
		\rho( \tilde{\mathfrak{L}}_{a_{2}} a_{1} ) \cdot f
	\right)
	+ 
	\frac{1}{2} 
	 \left(
		\rho \bigl( \tilde{\dop} \iop_{X_{1}} \xi_{2} \bigr) \cdot f
		- 
		\rho \bigl( \tilde{\dop} \iop_{X_{2}} \xi_{1} \bigr) \cdot f
	\right)  
	\notag\\
	& \quad
	+ \tilde{\rho}( [\xi_{1}, \xi_{2}]_{\tilde{L}} ) \cdot f
	+ 
	%\Bigl( 
		\tilde{\rho}( \mathcal{L}_{X_{1}} \xi_{2} ) \cdot f
		-
		\tilde{\rho}( \mathcal{L}_{X_{2}} \xi_{1} ) \cdot f
	%\Bigr)
	+
	%\left(
		\tilde{\rho}( \bar{\mathcal{L}}_{a_{1}} \xi_{2} ) \cdot f
		-
		\tilde{\rho}( \bar{\mathcal{L}}_{a_{2}} \xi_{1} ) \cdot f
	%\right)
	\notag\\
	&\qquad
	+
	\frac{1}{2}
	\Bigl(
		\tilde{\rho}( \mathfrak{L}_{a_{1}} a_{2} ) \cdot f
		-
		\tilde{\rho}( \mathfrak{L}_{a_{2}} a_{1} ) \cdot f
	\Bigr)
	-
	\frac{1}{2}
	\Bigl(
		\tilde{\rho} \bigl( \dop \iop_{X_{1}} \xi_{2} \bigr) \cdot f
		- \tilde{\rho} \bigl( \dop \iop_{X_{2}} \xi_{1}  \bigr) \cdot f
	\Bigr) 
	\notag\\
	&\quad
	+
	\frac{1}{2}
	\bar\rho( [a_{1}, a_{2}]_{\bar{L}} ) \cdot f
	+
	\frac{1}{2}
	\Bigl(
		\bar\rho( \bar{\mathcal{L}}_{a_{1}} a_{2} ) \cdot f
		-
		\bar\rho( \bar{\mathcal{L}}_{a_{2}} a_{1} ) \cdot f
	\Bigr)
	\notag\\
	&\qquad
	+
	%\left(
		\bar\rho( \mathcal{L}_{X_{1}} a_{2} ) \cdot f
		-
		\bar\rho( \mathcal{L}_{X_{2}} a_{1} ) \cdot f
	%\right)
	+
	%\left(
		\bar\rho( \tilde{\mathcal{L}}_{\xi_{1}} a_{2} ) \cdot f
		-
		\bar\rho( \tilde{\mathcal{L}}_{\xi_{2}} a_{1} ) \cdot f
	%\right) 
	\notag\\
	&\qquad
	+
	\frac{1}{2}
	\Bigl(
		\bar\rho( \bar{\mathfrak{L}}_{X_{1}} \xi_{2} ) \cdot f
		-
		\bar\rho( \bar{\mathfrak{L}}_{X_{2}} \xi_{1} ) \cdot f
	\Bigr)
	+
	\frac{1}{2}
	\Bigl(
		\bar\rho( \bar{\mathfrak{L}}_{\xi_{1}} X_{2} ) \cdot f
		-
		\bar\rho( \bar{\mathfrak{L}}_{\xi_{2}} X_{1} ) \cdot f
	\Bigr)
	\notag\\
	&\quad
	+ \boldsymbol\rho ( \iop_{\Xi_{2}}\iop_{\Xi_{1}} F ) \cdot f.
	\label{eq:Axiom2_act_left}
\end{align}
On the other hand, the second term in \eqref{eq:Axiom2_act} is expanded as 
\begin{align}
	[ \boldsymbol\rho(\Xi_{1}), \boldsymbol\rho(\Xi_{2}) ]  \cdot f
	&= 
	[ \rho(X_{1}), \rho(X_{2}) ]  \cdot f
	+ [ \rho(X_{1}), \tilde{\rho}(\xi_{2}) ]  \cdot f
	+ [ \rho(X_{1}), \bar{\rho}(a_{2}) ]  \cdot f
	\notag\\
	&\quad
	+ [ \tilde{\rho}(\xi_{1}), \rho(X_{2}) ]  \cdot f
	+ [ \tilde{\rho}(\xi_{1}), \tilde{\rho}(\xi_{2}) ]  \cdot f
	+ [ \tilde{\rho}(\xi_{1}), \bar{\rho}(a_{2}) ]  \cdot f
	\notag\\
	&\quad
	+ [ \bar{\rho}(a_{1}), \rho(X_{2}) ]  \cdot f
	+ [ \bar{\rho}(a_{1}), \tilde{\rho}(\xi_{2}) ]  \cdot f
	+ [ \bar{\rho}(a_{1}), \bar{\rho}(a_{2}) ]  \cdot f.
	\label{eq:Axiom2_act_right}
\end{align}
Using the fact that $(L, \rho, [\cdot, \cdot]_L)$, $(\tilde{L},
\tilde{\rho}, [\cdot, \cdot]_{\tilde{L}})$ and $(\bar{L}, \bar{\rho},
[\cdot, \cdot]_{F\text{-}\tilde{L}})$ are (twisted) Lie algebroids
, we obtain
\begin{align}
	&\boldsymbol\rho([\Xi_{1}, \Xi_{2}]_{F}) \cdot f - [ \boldsymbol\rho(\Xi_{1}), \boldsymbol\rho(\Xi_{2})] \cdot f
	\notag\\
	&=
	- \tilde{\iop}_{\xi_{2}} 
			\Bigl(
			\tilde{\mathcal{L}}_{\dop f} X_{1}
			-
			[ X_{1}, \tilde{\dop} f ]_{L}
			\Bigr)
	+\tilde{\iop}_{\xi_{1}} 
			\Bigl(
			\tilde{\mathcal{L}}_{\dop f} X_{2}
			-
			[ X_{2}, \tilde{\dop} f ]_{L}
			\Bigr)
	\notag\\
	&\quad
	+(\rho \tilde{\rho}^{\ast}+  \tilde{\rho}\rho^{\ast})
		\dop_{0}
		(\iop_{X_{1}} \xi_{2} - \iop_{X_{2}} \xi_{1}) 
		\cdot f
%	\notag\\
%	&\quad
	- \bar{\iop}_{a_{1}} \langle ( \tilde{\rho} \rho^{\ast} + \rho \tilde{\rho}^{\ast} ) \dop_0 f, \dop_0 a_{2} \rangle
	+
	\frac{1}{2} 
	\Bigl( 
		\rho\tilde{\rho}^{\ast} +  \tilde{\rho} \rho^{\ast}
	\Bigr)
	\dop_{0} \bar{\iop}_{a_{1}} a_{2}  \cdot f
	\notag\\
	&\quad
	+
	%\left(
	\frac{1}{2} \bar{\rho}\bar{\rho}^{\ast} \dop_{0} (\iop_{X_{1}} \xi_{2} + \iop_{X_{2}} \xi_{1}) \cdot f
	- \iop_{X_{1}} \langle \dop_{0} f, \bar{\rho} \bar{\rho}^{\ast} \dop_{0} \xi_{2} \rangle
	- \tilde{\iop}_{\xi_{1}} \langle \dop_{0} f, \bar{\rho} \bar{\rho}^{\ast} \dop_{0} X_{2} \rangle
%	\notag\\
%	&\quad\qquad
	%\right)
	\notag\\
	&\quad
	+
	\frac{1}{2}
	\bigl(
		\bar{\iop}_{a_{2}} \bar{\mathcal{L}}_{\bar{\dop} f} a_{1} 
		- \bar{\iop}_{a_{1}} \bar{\mathcal{L}}_{\bar{\dop} f} a_{2} 
	\bigr)
	+ \boldsymbol\rho ( \iop_{\Xi_{2}}\iop_{\Xi_{1}} F ) \cdot f
	.
	\label{eq:Axiom2_otsuri}
\end{align}
This does not vanish.
Therefore the quadruple $(\mathcal{V}, [\cdot, \cdot]_F, 
\boldsymbol{\rho}
,
\bbra{\cdot,\cdot})$ fails to satisfy Axiom C1 in general.
In order \eqref{eq:Axiom2_otsuri} to be zero, 
the following conditions are required;
\begin{align}
&\tilde{\mathcal{L}}_{\dop f} X_{1} - [ X_{1}, \tilde{\dop} f ]_{L} = 0,
\notag\\
& \tilde{\mathcal{L}}_{\dop f} X_{2} - [ X_{2}, \tilde{\dop} f ]_{L} = 0,
\notag\\
& \rho \tilde{\rho}^{\ast} + \tilde{\rho}\rho^{\ast} = 0.
\label{eq:AxiomC1_condition1}
\end{align}
We also have the conditions
\begin{align}
&
\bar{\iop}_{a_{2}} \bar{\mathcal{L}}_{\bar{\dop} f} a_{1} 
- \bar{\iop}_{a_{1}} \bar{\mathcal{L}}_{\bar{\dop} f} a_{2}  = 0,
\notag \\
&
\bar{\rho} \bar{\rho}^{\ast}  = 0,
\qquad
\boldsymbol\rho ( \iop_{\Xi_{2}}\iop_{\Xi_{1}} F ) \cdot f = 0.
\label{eq:AxiomC1_condition2}
\end{align}
The conditions \eqref{eq:AxiomC1_condition1} are nothing but the ones 
that $(L, \tilde{L})$ becomes a Lie bialgebroid \cite{LiWeXu}.
The last condition in \eqref{eq:AxiomC1_condition2} is a sophisticated expression of 
the second equation in \eqref{eq:condition_F_1}.
Indeed, when we employ $\bar{\rho} = \mathrm{id}$, 
the last equation in \eqref{eq:AxiomC1_condition2} becomes equivalent to the one in \eqref{eq:condition_F_1}.
We will comment on \eqref{eq:AxiomC1_condition1} and
\eqref{eq:AxiomC1_condition2} in due course.

\paragraph{Proof of Axiom C2}
We next study Axiom C2.
The left-hand side in \eqref{def:Axiom4} is expanded 
and rewritten as
\begin{align}
\bbra{\mathcal{D}f, \mathcal{D}g} 
&=
\bbra{(\dop + \tilde{\dop} + \bar{\dop})f, (\dop + \tilde{\dop} + \bar{\dop})g}  \notag\\
&=
\frac{1}{2} \bigr\{ 
\iop_{\dop g} \tilde{\dop} f
+ \iop_{\tilde{\dop} g} \dop f
+ \iop_{\bar{\dop} g} \bar{\dop} f
\bigl\}
\notag\\
&=
\frac{1}{2} \bigr\{ 
\tilde{\rho} ( \dop g ) \cdot f
+ \rho ( \tilde{\dop} g ) \cdot f
+ \bar{\rho} ( \bar\dop g ) \cdot f
\bigl\}
\notag \\
&=
\frac{1}{2} \bigr\{ 
\tilde{\rho} \rho^{\ast} \dop_{0}  g \cdot f
+ \rho \tilde{\rho}^{\ast} \dop_{0} g \cdot f
+ \bar{\rho} \bar{\rho}^{\ast} \dop_{0} g \cdot f
\bigl\}
\notag\\
&=
\frac{1}{2} \bigr\{ 
(
\tilde{\rho} \rho^{\ast} 
+ \rho \tilde{\rho}^{\ast}
+ \bar{\rho} \bar{\rho}^{\ast}
)
( \dop_{0} g \cdot f)
\bigl\}.
\end{align}
This again does not vanish in general.
In order this to be zero, we have the condition;
\begin{align}
\tilde{\rho} \rho^{\ast} 
+ \rho \tilde{\rho}^{\ast}
+ \bar{\rho} \bar{\rho}^{\ast} = 0.
\label{eq:AxiomC2_condition}
\end{align}
We note that this condition is automatically satisfied when 
\eqref{eq:AxiomC1_condition1} and \eqref{eq:AxiomC1_condition2} are satisfied.
If we keep only the $O(D,D)$ part by dropping out the gauge sector,
the skew-symmetric property of the anchor $\rho,\tilde{\rho}$ is
naturally recovered \cite{LiWeXu}.
We also note that the situation is consistent with the fact that Axiom
C1 actually implies Axiom C2 in Courant algebroids \cite{Uchino}.

\paragraph{Proof of Axiom C3}
Axiom C3 is the modified Jacobi identity for the twisted C-bracket;
\begin{align}
	[[\Xi_{1}, \Xi_{2} ]_{F},\Xi_{3} ]_{F} + \mathrm{c.p.} = \mathcal{D} T_{F} ( \Xi_{1}, \Xi_{2}, \Xi_{3}),
	\label{eq:Axiom1}
\end{align}
where $T_{F}$ is defined by \eqref{eq:Lemma3.2_Fbra}.
In the following, we calculate the left-hand side of \eqref{eq:Axiom1},
and examine the conditions for Axiom C3.
The left-hand side of \eqref{eq:Axiom1} is decomposed as
\begin{align}
	[[\Xi_{1}, \Xi_{2} ]_{F},\Xi_{3} ]_{F} + \mathrm{c.p.} =
 \mathrm{Jac}_{F} (\Xi_{1}, \Xi_{2}, \Xi_{3})
 = I'_{1} + I'_{2} + I'_{3},
\end{align}
where $I'_{1}, I'_{2}, I'_{3}$ are $\Gamma (\tilde{L})$, $\Gamma (L)$
and $\Gamma (\bar{L})$ parts in $\mathrm{Jac}_F (\Xi_{1}, \Xi_{2}, \Xi_{3})$, respectively.
The explicit forms of $I'_{1}, I'_{2}, I'_{3}$ are given in Appendix \ref{Appendix:AxiomC3}.

It is useful to evaluate $\mathrm{Jac}_F (\Xi_{1}, \Xi_{2},\Xi_{3})$ in powers of $F^M {}_{NP}$.
As is obvious in the right-hand side in 
\eqref{eq:Lemma3.2_Fbra}, 
we expect that the $\mathcal{O} (F^2)$ terms in $\mathrm{Jac}_F
(\Xi_{1}, \Xi_{2},\Xi_{3})$ must vanish.
Indeed, the $\mathcal{O}(F^2)$ terms in $\mathrm{Jac}_F (\Xi_{1},
\Xi_{2},\Xi_{3})$ are calculated as
\begin{align}
	\mathrm{Jac}_{F} (\Xi_{1}, \Xi_{2}, \Xi_{3}) |_{\mathcal{O}(F^{2})}
	&=
	\bold{i}_{\Xi_{3}}\bold{i}_{(\bold{i}_{\Xi_{2}} \bold{i}_{\Xi_{1}F})}F + \mathrm{c.p.}
	=
	\Xi_{3}^{R} \Xi_{2}^{K} \Xi_{1}^{N} F^{M}{}_{P[R}  F^{P}{}_{NK]}.
	\label{eq:2nd_order_f}
\end{align}
This vanishes due to the Jacobi identity of $F^M {}_{NP}$.

We next evaluate the $\mathcal{O} (F^1)$ terms in $\mathrm{Jac}_F (\Xi_{1},
\Xi_{2},\Xi_{3})$. 
Since the $I'_{1}$ and $I'_{2}$ parts have essentially the same
structures and they interchange by $X \leftrightarrow \xi$,
we focus only on $I'_{1}$ and $I'_{3}$ parts.
Firstly, we have 
\begin{align}
	&
	I'_{1} |_{\mathcal{O}(F)}
	\notag\\
	&=
	[\bold{i}_{\Xi_{2}} \bold{i}_{\Xi_{1}}F_{\mu} \tilde{\partial}^{\mu}, \xi_{3}]_{\tilde{L}}
	+ 
	\mathcal{L}_{(\bold{i}_{\Xi_{2}} \bold{i}_{\Xi_{1}}F^{\mu} \partial_{\mu})} \xi_{3}
	-
	\mathcal{L}_{X_{3}} (\bold{i}_{\Xi_{2}} \bold{i}_{\Xi_{1}}F_{\mu} \tilde{\partial}^{\mu})
	+ 
	\bar{\mathcal{L}}_{(\bold{i}_{\Xi_{2}} \bold{i}_{\Xi_{1}}F_{\alpha} \bar{\partial}^{\alpha})} \xi_{3}
	- 
	\bar{\mathcal{L}}_{a_{3}} (\bold{i}_{\Xi_{2}} \bold{i}_{\Xi_{1}}F_{\mu} \tilde{\partial}^{\mu})
	\notag\\
	&\quad
	+
	\frac{1}{2} \mathfrak{L}_{(\bold{i}_{\Xi_{2}} \bold{i}_{\Xi_{1}}F_{\alpha} \bar{\partial}^{\alpha})} a_{3}
	- 
	\frac{1}{2} \mathfrak{L}_{a_{3}} (\bold{i}_{\Xi_{2}} \bold{i}_{\Xi_{1}}F_{\alpha} \bar{\partial}^{\alpha})
	-
	\frac{1}{2} \dop \iop_{(\bold{i}_{\Xi_{2}} \bold{i}_{\Xi_{1}}F^{\mu} \partial_{\mu})} \xi_{3}
	+
	\frac{1}{2} \iop_{X_{3}} (\bold{i}_{\Xi_{2}} \bold{i}_{\Xi_{1}}F)
	+
	\bold{i}_{\Xi_{3}} \bold{i}_{[\Xi_{1}, \Xi_{2} ]_{C}} F
	+ 
	\mathrm{c.p.}
	\notag\\
%%%%%
	&=
	\frac{1}{2} \dop ( \bold{i}_{\Xi_{3}} \bold{i}_{\Xi_{2}} \bold{i}_{\Xi_{1}}F)
	\notag\\
	&
	\quad
	+
	\bigl\{
	\Xi_{2}^{Q}\Xi_{1}^{P} F^{K}{}_{PQ} (\partial_{K} \xi_{\mu 3}) \tilde{\partial}^{\mu}
	- 
	\frac{1}{2} \Xi_{3}^{K} \Xi^{P}_{1}  F_{\mu}{}_{NK} (\partial^{N} \Xi_{2P}) \tilde{\partial}^{\mu}
	+ 
	\frac{1}{2} \Xi_{3}^{K} \Xi^{P}_{2} F_{\mu}{}_{NK} (\partial^{N} \Xi_{1P}) \tilde{\partial}^{\mu} 
	+ 
	\mathrm{c.p.}
	\bigr\}.
\label{eq:AxiomC3_eq1}
\end{align}
Here $[\Xi_{1}, \Xi_{2} ]_{C}$ is the C-bracket without the term
$\bold{i}_{\Xi_1} \bold{i}_{\Xi_2} F$.
We have also introduced the notations
\begin{align}
\bold{i}_{\Xi_2} \bold{i}_{\Xi_1} F_{\mu} \tilde{\del}^{\mu} = \Xi_2^M
 \Xi_1^N F_{\mu MN} \tilde{\del}^{\mu},
\quad
\bold{i}_{\Xi_2} \bold{i}_{\Xi_1} F^{\mu} {\del}_{\mu} = \Xi_2^M
 \Xi_1^N F^{\mu} {}_{MN} {\del}_{\mu},
\quad
\bold{i}_{\Xi_2} \bold{i}_{\Xi_1} F_{\alpha} \bar{\del}^{\alpha} = \Xi_2^M
 \Xi_1^N F_{\alpha MN} \bar{\del}^{\alpha}.
\end{align}
\footnote{
Although the non-zero components of $F$ are only $F_{\alpha} {}^{\beta
\gamma}$, we have formally kept all the components of $F$ in order for general discussion.
}.
The first term in the right-hand side 
in the last expression of \eqref{eq:AxiomC3_eq1}
contributes to $\mathcal{D} T_{F}$
while the remaining terms must vanish.
This therefore implies the condition;
\begin{align}
	F^{K}{}_{PQ} \partial_{K} \ast = 0
\end{align}
for any quantities $*$.

The last is the $\mathcal{O} (F^0)$ terms.
We evaluate each term that contains specific quantities.
For example, terms that contain one $a$ and two $\xi$s are given by 
\begin{align}
	I'_{1} |_{\mathcal{O}(F^{0}),\,a \xi \xi}
	&=
	[\bar{\mathcal{L}}_{a_{1}} \xi_{2}, \xi_{3}]_{\tilde{L}}
	-
	[\bar{\mathcal{L}}_{a_{2}} \xi_{1}, \xi_{3}]_{\tilde{L}}
	+
	\bar{\mathcal{L}}_{\tilde{\mathcal{L}}_{\xi_{1}} a_{2}} \xi_{3}
	-
	\bar{\mathcal{L}}_{\tilde{\mathcal{L}}_{\xi_{2}} a_{1}} \xi_{3}
	-
	\bar{\mathcal{L}}_{a_{3}} [\xi_{1}, \xi_{2}]_{\tilde{L}}
	+
	\mathrm{c.p.}
	\label{eq:aaxi}
\end{align}
By using the following identity 
\begin{align}
	[\bar{\mathcal{L}}_{a_{1}} \xi_{2}, \xi_{3}]_{\tilde{L}}
	-
	[\bar{\mathcal{L}}_{a_{1}} \xi_{3}, \xi_{2}]_{\tilde{L}}
	+
	\bar{\mathcal{L}}_{\tilde{\mathcal{L}}_{\xi_{3}} a_{1}} \xi_{2}
	-
	\bar{\mathcal{L}}_{\tilde{\mathcal{L}}_{\xi_{2}} a_{1}} \xi_{3}
	-
	\bar{\mathcal{L}}_{a_{1}} [\xi_{2}, \xi_{3}]_{\tilde{L}}
	=
	0,
\end{align}
we find $I'_{1} |_{\mathcal{O}(F^0), a \xi \xi} = 0$.
Terms that contain one $X$ and two $\xi$s are given by
\begin{align}
	I'_{1} |_{\mathcal{O}(F^0),X \xi \xi}
	&=
	\frac{1}{2} \bar{\iop}_{(\iop_{X_{1}} \bar{\dop} \xi_{2})} \bar{\dop} \xi_{3}
	-
	\frac{1}{2} \bar{\iop}_{(\iop_{X_{2}} \bar{\dop} \xi_{1})} \bar{\dop} \xi_{3}
	+
	\frac{1}{2} \bar{\iop}_{(\tilde{\iop}_{\xi_{1}} \bar{\dop} X_{2})} \bar{\dop} \xi_{3}
	-
	\frac{1}{2} \bar{\iop}_{(\tilde{\iop}_{\xi_{2}} \bar{\dop} X_{1})} \bar{\dop} \xi_{3}
	+
	\mathrm{c.p.}
\end{align}
This neither contributes to $\mathcal{D} T_F$ nor vanishes in general.
Terms that contain only $a$ becomes
\begin{align}
	I'_{1} |_{\mathcal{O}(F^0),a a a}
	&=
	- 
	\frac{1}{2} \bar{\iop}_{a_{3}} \bar{\dop} \bar{\iop}_{a_{1}} \dop  a_{2}
	+ 
	\frac{1}{2} \bar{\iop}_{a_{3}} \bar{\dop} \bar{\iop}_{a_{2}} \dop  a_{1}
	+
	\frac{1}{4} \iop_{(\bar{\iop}_{a_{1}}  \bar{\dop} a_{2})} \dop a_{3}
	-
	\frac{1}{4} \iop_{(\bar{\iop}_{a_{2}}  \bar{\dop} a_{1})} \dop a_{3}
	\notag\\
	&\quad
	- 
	\frac{1}{4} \bar{\iop}_{a_{3}} \dop  \bar{\iop}_{a_{1}} \bar{\dop} a_{2}
	+ 
	\frac{1}{4} \bar{\iop}_{a_{3}} \dop  \bar{\iop}_{a_{2}} \bar{\dop} a_{1}
	+
	\frac{1}{4} \mathfrak{L}_{[a_{1}, a_{2}]_{\bar{L}}} a_{3}
	- 
	\frac{1}{4} \mathfrak{L}_{a_{3}} [a_{1}, a_{2}]_{\bar{L}}
	+
	\mathrm{c.p.}
	\notag\\
	&=
	 \frac{1}{4} \dop \bar{\iop}_{a_{2}} \bar{\iop}_{a_{3}} \bar{\dop} a_{1} + \mathrm{c.p.}
\end{align}
Here we have expanded the Lie(-like) derivatives and collected all the
$\mathrm{c.p.}$ parts together.
Similar treatments have been done for the other parts and we finally obtain
the following result;
\begin{align}
	I'_{1} |_{\mathcal{O}(\mathcal{F}^0)}
	&=
	\dop T(e_{1}, e_{2}, e_{3})
	\notag\\
	&\quad
	+ \frac{1}{4} \dop
		\Bigl\{
		\bigl(
		\tilde{\iop}_{\xi_{2}} \bar{\iop}_{a_{3}}  \bar{\dop} X_{1}
		+ \iop_{X_{2}} \bar{\iop}_{a_{3}} \bar{\dop} \xi_{1}
		+ \bar{\iop}_{a_{2}} \iop_{X_{3}} \dop a_{1}
		+ \bar{\iop}_{a_{2}} \tilde{\iop}_{\xi_{3}} \tilde{\dop} a_{1}
		+ \bar{\iop}_{a_{2}} \bar{\iop}_{a_{3}} \bar{\dop} a_{1}
		\bigr) 
	\notag\\
	&\qquad\quad
		- 
		\bigl(
		\tilde{\iop}_{\xi_{3}} \bar{\iop}_{a_{2}} \bar{\dop} X_{1}
		+ \iop_{X_{3}} \bar{\iop}_{a_{2}} \bar{\dop} \xi_{1}
		+ \bar{\iop}_{a_{3}} \iop_{X_{2}} \dop a_{1}
		+ \bar{\iop}_{a_{3}} \tilde{\iop}_{\xi_{2}} \tilde{\dop} a_{1}
		\bigr)
		+ \mathrm{c.p.}
		 \Bigr\}
	\notag\\
%%%%%
	&\quad
	+
	\Bigl\{
	-
	\iop_{X_{3}} \dop [\xi_{1}, \xi_{2}]_{\tilde{L}}
	+
	\iop_{X_{3}} \tilde{\mathcal{L}}_{\xi_{1}} \dop \xi_{2}
	-
	\iop_{X_{3}} \tilde{\mathcal{L}}_{\xi_{2}} \dop \xi_{1}
	\notag\\
	&\quad\qquad
	-
	\frac{1}{2} \mathcal{L}_{\tilde{\dop} \tilde{\iop}_{\xi_{1}} X_{2}} \xi_{3}
	+
	\frac{1}{2} \mathcal{L}_{\tilde{\dop} \tilde{\iop}_{\xi_{2}} X_{1}} \xi_{3}
	-
	\frac{1}{2} [\dop \tilde{\iop}_{\xi_{1}} X_{2}, \xi_{3}]_{\tilde{L}}
	+
	\frac{1}{2} [\dop \tilde{\iop}_{\xi_{2}} X_{1}, \xi_{3}]_{\tilde{L}}
	+
	\mathrm{c.p.}
	\Bigr\}
	\notag\\
%%%%%%
	&\quad
	+
	\biggl\{
	\frac{1}{2} \iop_{\iop_{X_{1}} \bar{\dop} \xi_{2}} \bar{\dop} \xi_{3}
	-
	\frac{1}{2} \iop_{\iop_{X_{2}} \bar{\dop} \xi_{1}} \bar{\dop} \xi_{3}
	+
	\frac{1}{2} \iop_{\tilde{\iop}_{\xi_{1}} \bar{\dop} X_{2}} \bar{\dop} \xi_{3}
	-
	\frac{1}{2} \iop_{\tilde{\iop}_{\xi_{2}} \bar{\dop} X_{1}} \bar{\dop} \xi_{3}
	\notag\\
	&\quad\qquad
	+
	\frac{1}{2}
	[\mathfrak{L}_{a_{1}} a_{2}, \xi_{3}]_{\bar{L}}
	-
	\frac{1}{2}
	[\mathfrak{L}_{a_{1}} a_{2}, \xi_{3}]_{\bar{L}}
	+
	\frac{1}{2}
	\mathcal{L}_{\tilde{\mathfrak{L}}_{a_{1}} a_{2}} \xi_{3}
	-
	\frac{1}{2}
	\mathcal{L}_{\tilde{\mathfrak{L}}_{a_{1}} a_{2}} \xi_{3}
	\notag\\
	&\quad\qquad
	+
	\frac{1}{2}
	\mathfrak{L}_{\tilde{\mathcal{L}}_{\xi_{3}} a_{1}} a_{2}
	-
	\frac{1}{2}
	\mathfrak{L}_{\tilde{\mathcal{L}}_{\xi_{3}} a_{2}} a_{1}
	+
	\frac{1}{2}
	\mathfrak{L}_{a_{1}} \tilde{\mathcal{L}}_{\xi_{3}} a_{2}
	-
	\frac{1}{2}
	\mathfrak{L}_{a_{2}} \tilde{\mathcal{L}}_{\xi_{3}} a_{1}
	\notag\\
	&\quad\qquad
	-
	\frac{1}{2} \bar{\mathcal{L}}_{a_{1}} \bar{\mathcal{L}}_{a_{2}} \xi_{3}
	+
	\frac{1}{2} \bar{\mathcal{L}}_{a_{2}} \bar{\mathcal{L}}_{a_{1}} \xi_{3}
	+
	\frac{1}{2} \bar{\mathcal{L}}_{\bar{\mathcal{L}}_{a_{1}} a_{2}} \xi_{3}
	-
	\frac{1}{2} \bar{\mathcal{L}}_{\bar{\mathcal{L}}_{a_{2}} a_{1}} \xi_{3}
	+
	\mathrm{c.p.}
	\biggr\}.
\end{align}
The first, the second and the third terms precisely give $\dop T_{F}
|_{\mathcal{O} (F^0)}$.
After expanding the fourth to the sixth terms and collecting all
together, we find
\begin{align}
	I'_{1} |_{\mathcal{O}(F^0)}
	&=
	\dop T_{F} (\Xi_{1}, \Xi_{2}, \Xi_{3}) |_{\mathcal{O} (F^0)}
	\notag\\
	&\quad
	-
	\frac{1}{2} \eta_{KL} \Xi_{3}^{K} (\partial^{P} \Xi_{1}{}^{L}) (\partial_{P} \xi_{2 \mu}) \partial^{\mu}
	+
	\frac{1}{2} \eta_{KL} \Xi_{2}^{K} (\partial^{P} \Xi_{1}{}^{L}) (\partial_{P} \xi_{3 \mu}) \partial^{\mu}.
\end{align}
In order Axiom C3 holds, we therefore impose the condition $\eta^{MN}
\del_{M} \ast \partial_{N} \xi = 0$.
In summary, we have the following conditions in $I'_1$ for Axiom C3;
\begin{align}
F^{K}{}_{PQ} \partial_{K} \ast = 0,
\qquad
\eta^{MN} \del_M \ast \partial_{N} \xi^{\mu} = 0.
\end{align}
Similar calculations in $I'_2$ lead to the conditions 
\begin{align}
F^{K}{}_{PQ} \partial_{K} \ast = 0,
\qquad
\eta^{MN} \del_M \ast \partial_{N} X_{\mu} = 0.
\end{align}

Finally, we evaluate the remaining part $I'_3$.
The total expression of $I'_3$ is given by \eqref{eq:I3prime} in Appendix.
The $\mathcal{O} (F^1)$ in $I'_3$ is evaluated as 
\begin{align}
	I'_3|_{\mathcal{O}(F)}
	&=
	\frac{1}{2} [\bold{i}_{\Xi_{2}} \bold{i}_{\Xi_{1}}F_{\alpha} \bar{\partial}^{\alpha}, a_{3}]_{\bar{L}}
	+ 
	\frac{1}{2} \bar{\mathcal{L}}_{(\bold{i}_{\Xi_{2}} \bold{i}_{\Xi_{1}}F_{\alpha} \bar{\partial}^{\alpha})} a_{3}
	-
	\frac{1}{2} \bar{\mathcal{L}}_{a_{3}} (\bold{i}_{\Xi_{2}} \bold{i}_{\Xi_{1}}F_{\alpha} \bar{\partial}^{\alpha})
	\notag\\
	&\quad 
	+ 
	\mathcal{L}_{(\bold{i}_{\Xi_{2}} \bold{i}_{\Xi_{1}}F^{\mu} \partial_{\mu})} a_{3}
	-
	\mathcal{L}_{X_{3}} (\bold{i}_{\Xi_{2}} \bold{i}_{\Xi_{1}}F_{\alpha} \bar{\partial}^{\alpha})
	+ 
	\tilde{\mathcal{L}}_{(\bold{i}_{\Xi_{2}} \bold{i}_{\Xi_{1}}F_{\mu} \tilde{\partial}^{\mu})} a_{3}
	-
	\tilde{\mathcal{L}}_{\xi_{3}} (\bold{i}_{\Xi_{2}} \bold{i}_{\Xi_{1}}F_{\alpha} \bar{\partial}^{\alpha})
	\notag\\
	&\quad 
	+
	\frac{1}{2} \bar{\mathfrak{L}}_{(\bold{i}_{\Xi_{2}} \bold{i}_{\Xi_{1}}F^{\mu} \partial_{\mu})} \xi_{3}
	- 
	\frac{1}{2} \bar{\mathfrak{L}}_{X_{3}} (\bold{i}_{\Xi_{2}} \bold{i}_{\Xi_{1}}F_{\mu} \tilde{\partial}^{\mu})
	+
	\frac{1}{2} \bar{\mathfrak{L}}_{(\bold{i}_{\Xi_{2}} \bold{i}_{\Xi_{1}}F^{\mu} \partial_{\mu})} \xi_{3}
	\notag\\
	&\quad 
	-
	\frac{1}{2} \bar{\mathfrak{L}}_{\xi_{3}} (\bold{i}_{\Xi_{2}} \bold{i}_{\Xi_{1}}F^{\mu} \partial_{\mu})
	+
	\bold{i}_{\Xi_{3}} \bold{i}_{[\Xi_{1}, \Xi_{2} ]_{C}} F_{\alpha} \bar{\partial}^{\alpha}
	+ \mathrm{c.p.}	
	\notag\\
	&=
	\frac{1}{2} \bar{\dop} (\bold{i}_{\Xi_{3}} \bold{i}_{\Xi_{2}} \bold{i}_{\Xi_{1}}F)
	\notag\\
	&
	\quad
	+
	\left\{
	\Xi_{2}^{Q}\Xi_{1}^{P} F^{K}{}_{PQ} (\partial_{K} a_{3 \alpha}) \bar{\partial}^{\alpha}
	- 
	\frac{1}{2} \Xi_{3}^{K} \Xi^{P}_{1}  F_{\alpha}{}_{NK} (\partial^{N} \Xi_{2P}) \bar{\partial}^{\alpha}
	+ 
	\frac{1}{2} \Xi_{3}^{K} \Xi^{P}_{2} F_{\alpha}{}_{NK} (\partial^{N} \Xi_{1P}) \bar{\partial}^{\alpha} 
	+ 
	\mathrm{c.p.}
	\right\}.
\end{align}
As in the case of $I_{1}'$, we find that the condition $F^{K}{}_{PQ}
\partial_{K} \ast = 0$ is necessary for Axiom C3.

We next examine the $\mathcal{O} (F^0)$ contributions.
Since we have many terms in $I'_3|_{\mathcal{O} (F^0)}$ we pick up them
one by one. For example, the terms including three $a$s are given by
\begin{align}
	I'_{3}|_{\mathcal{O}(F^{0}),\:aaa}
	&=
	\frac{1}{4} [[a_{1}, a_{2}]_{\bar{L}}, a_{3}]_{\bar{L}}
	+
	\frac{1}{4} [\bar{\mathcal{L}}_{a_{1}} a_{2}, a_{3}]_{\bar{L}}
	-
	\frac{1}{4} [\bar{\mathcal{L}}_{a_{2}} a_{1}, a_{3}]_{\bar{L}}
	-
	\frac{1}{4} \bar{\mathcal{L}}_{a_{3}} [a_{1}, a_{2}]_{\bar{L}}
	\notag\\
	&\quad
	+
	\frac{1}{4} \bar{\mathcal{L}}_{[a_{1}, a_{2}]_{\bar{L}}} a_{3}
	+
	\frac{1}{4} \bar{\mathcal{L}}_{\bar{\mathcal{L}}_{a_{1}} a_{2}} a_{3}
	-
	\frac{1}{4} \bar{\mathcal{L}}_{\bar{\mathcal{L}}_{a_{2}} a_{1}} a_{3}
	\notag\\
	&\quad
	-
	\frac{1}{4} \bar{\mathcal{L}}_{a_{3}} \bar{\mathcal{L}}_{a_{1}} a_{2}
	+
	\frac{1}{4} \bar{\mathcal{L}}_{a_{3}} \bar{\mathcal{L}}_{a_{2}} a_{1}
	+ 
	\frac{1}{2} \mathcal{L}_{\tilde{\mathfrak{L}}_{a_{1}} a_{2}} a_{3}
	-
	\frac{1}{2} \mathcal{L}_{\tilde{\mathfrak{L}}_{a_{2}} a_{1}} a_{3}
	\notag\\
	&\quad
	+
	\frac{1}{2} \tilde{\mathcal{L}}_{\mathfrak{L}_{a_{1}} a_{2}} a_{3}
	-
	\frac{1}{2} \tilde{\mathcal{L}}_{\mathfrak{L}_{a_{2}} a_{1}} a_{3}
	+
	\mathrm{c.p.}
	\notag\\
%%%%%
	&=
	\frac{1}{4} [\bar{\iop}_{a_{1}} \bar{\dop} a_{2}, a_{3}]_{\bar{L}}
	-
	\frac{1}{4} [\bar{\iop}_{a_{2}} \bar{\dop} a_{1}, a_{3}]_{\bar{L}}
	+
	\frac{1}{4} \bar{\iop}_{[a_{1}, a_{2}]_{\bar{L}}} \bar{\dop} a_{3}
	-
	\frac{1}{4} \bar{\iop}_{a_{3}} \bar{\dop} [a_{1}, a_{2}]_{\bar{L}}
	\notag\\
	&\quad
	+
	\frac{1}{4} \bar{\iop}_{(\bar{\iop}_{a_{1}} \bar{\dop} a_{2})} \bar{\dop} a_{3}
	-
	\frac{1}{4} \bar{\iop}_{(\bar{\iop}_{a_{2}} \bar{\dop} a_{1})} \bar{\dop} a_{3}
	-
	\frac{1}{4} \bar{\iop}_{a_{3}} \bar{\dop} \bar{\iop}_{a_{1}} \bar{\dop} a_{2}
	-
	\frac{1}{4} \bar{\iop}_{a_{3}} \bar{\dop} \bar{\iop}_{a_{2}} \bar{\dop} a_{1}
	\notag\\
	&\quad
	+
	\frac{1}{2} \iop_{(\bar{\iop}_{a_{1}} \tilde{\dop} a_{2})} \dop a_{3}
	-
	\frac{1}{2} \iop_{(\bar{\iop}_{a_{2}} \tilde{\dop} a_{1})} \dop a_{3}
	+
	\frac{1}{2} \tilde{\iop}_{(\bar{\iop}_{a_{1}} \dop a_{2})} \tilde{\dop} a_{3}
	-
	\frac{1}{2} \tilde{\iop}_{(\bar{\iop}_{a_{2}} \dop a_{1})} \tilde{\dop} a_{3}
	+
	\mathrm{c.p.}
\end{align}
Here we have decomposed the Lie(-like) derivatives into the interior products and the derivative operators.
It is easy to show that the first and the second lines 
together with the c.p. parts in the last expression contribute to $T_F$.

We find that the terms including two $\xi$s and one $a$ vanish;
\begin{align}
	I'_{3}|_{\mathcal{O}(F^{0}),\:\xi \xi a}
	&=
	\tilde{\mathcal{L}}_{[\xi_{1}, \xi_{2}]_{\tilde{L}}} a_{3}
	-
	\tilde{\mathcal{L}}_{\xi_{3}} \tilde{\mathcal{L}}_{\xi_{1}} a_{2}
	+
	\tilde{\mathcal{L}}_{\xi_{3}} \tilde{\mathcal{L}}_{\xi_{2}} a_{1}
	+
	\mathrm{c.p.} = 0.
\end{align}
Here we have used the identity
\begin{align}
	\tilde{\mathcal{L}}_{[\xi_{1}, \xi_{2}]_{\tilde{L}}} a_{3}
	-
	\tilde{\mathcal{L}}_{\xi_{1}} \tilde{\mathcal{L}}_{\xi_{2}} a_{3}
	+
	\tilde{\mathcal{L}}_{\xi_{2}} \tilde{\mathcal{L}}_{\xi_{1}} a_{3}
	= 0.
\end{align}

We also find that the $\xi \xi X$ terms give a total derivative term;
\begin{align}
	I'_{3}|_{\mathcal{O}(F^{0}),\:\xi \xi X}
	&=
	-
	\frac{1}{2} \tilde{\mathcal{L}}_{\xi_{3}} \bar{\mathfrak{L}}_{X_{1}} \xi_{2}
	+
	\frac{1}{2} \tilde{\mathcal{L}}_{\xi_{3}} \bar{\mathfrak{L}}_{X_{2}} \xi_{1}
	-
	\frac{1}{2} \tilde{\mathcal{L}}_{\xi_{3}} \bar{\mathfrak{L}}_{\xi_{1}} X_{2}
	+
	\frac{1}{2} \tilde{\mathcal{L}}_{\xi_{3}} \bar{\mathfrak{L}}_{\xi_{2}} X_{1}
	\notag\\
	&\quad
	+ 
	\frac{1}{2} \bar{\mathfrak{L}}_{\tilde{\mathcal{L}}_{\xi_{1}} X_{2}} \xi_{3}
	-
	\frac{1}{2} \bar{\mathfrak{L}}_{\tilde{\mathcal{L}}_{\xi_{2}} X_{1}} \xi_{3}
	+ 
	\frac{1}{4} \bar{\mathfrak{L}}_{\tilde{\dop} \iop_{X_{1}} \xi_{2}} \xi_{3}
	-
	\frac{1}{4} \bar{\mathfrak{L}}_{\tilde{\dop} \iop_{X_{2}} \xi_{1}} \xi_{3}
	- 
	\frac{1}{2} \bar{\mathfrak{L}}_{X_{3}} [\xi_{1}, \xi_{2}]_{\tilde{L}}
	\notag\\
	&\quad
	+
	\frac{1}{2} \bar{\mathfrak{L}}_{[\xi_{1}, \xi_{2}]_{\tilde{L}}} X_{3}
	-
	\frac{1}{2} \bar{\mathfrak{L}}_{\xi_{3}} \tilde{\mathcal{L}}_{\xi_{1}} X_{2}
	+
	\frac{1}{2} \bar{\mathfrak{L}}_{\xi_{3}} \tilde{\mathcal{L}}_{\xi_{2}} X_{1}
	-
	\frac{1}{4} \bar{\mathfrak{L}}_{\xi_{3}} \tilde{\dop} \iop_{X_{1}} \xi_{2} 
	+
	\frac{1}{4} \bar{\mathfrak{L}}_{\xi_{3}} \tilde{\dop} \iop_{X_{2}} \xi_{1}
	+
	\mathrm{c.p.}
	\notag\\
	&=
	\frac{1}{2} 
	\bar{\dop}
	\Bigl(
		\tilde{\iop}_{[\xi_{1}, \xi_{2}]_{L}} X_{3}
		- 
		\frac{1}{2}\tilde{\iop}_{\xi_{3}} \tilde{\dop} \tilde{\iop}_{\xi_{1}}X_{2}
		+ 
		\frac{1}{2}\tilde{\iop}_{\xi_{3}} \tilde{\dop} \tilde{\iop}_{\xi_{2}}X_{1}
	\Bigr)
	+
	\mathrm{c.p.}
\end{align}
The $XX\xi$ terms have similar expression and we have 
\begin{align}
	I'_{3}|_{\mathcal{O}(F^{0}),\:\xi \xi X}
	+
	I'_{3}|_{\mathcal{O}(F^{0}),\:X X \xi}
	&=
	\bar{\dop} T({e_{1}, e_{2}}, e_{3}).
\end{align}
The other terms are also evaluated similarly.
We then finally obtain the following result;
\begin{align}
	I'_{3} |_{\mathcal{O} (F^0)}
	&=
	\dop T_{F} (\Xi_{1}, \Xi_{2}, \Xi_{3}) |_{\mathcal{O} (F^0)}
	\notag\\
	&\quad
	-
	\frac{1}{2} \eta_{KL} \Xi_{3}^{K} (\partial^{P} \Xi_{1}{}^{L}) (\partial_{P} a_{2 \alpha}) \bar{\partial}^{\mu}
	+
	\frac{1}{2} \eta_{KL} \Xi_{2}^{K} (\partial^{P} \Xi_{1}{}^{L}) (\partial_{P} a_{3 \alpha}) \bar{\partial}^{\mu}.
\end{align}
Therefore, we find the following conditions from $I'_3$;
\begin{align}
F^{K}{}_{PQ} \partial_{K} \ast = 0,
\qquad
\eta^{MN} \del_M \ast \partial_{N} a_{\alpha}= 0.
\end{align}

Collecting all the conditions from $I'_{1}, I'_{2}, I'_{3}$
together, we conclude that the necessary conditions for Axiom C3
are given by
\begin{align}
F^{K}{}_{PQ} \partial_{K} \ast = 0,
\qquad
\eta^{MN} \del_M \ast \partial_{N} \ast= 0.
\label{eq:Courant_conditions}
\end{align}
They are nothing but the physical conditions
\eqref{eq:physical_conditions} in gauged DFT.
We note that the conditions \eqref{eq:AxiomC1_condition1} and
\eqref{eq:AxiomC1_condition2} for Axiom C1 and hence
\eqref{eq:AxiomC2_condition} for Axiom C2 
always hold provided that \eqref{eq:Courant_conditions} are satisfied.

This result reveals that the mathematical origin of the physical conditions in
gauged DFT is the compatibility conditions for three algebroids to compose
a Courant algebroid. 
This is a generalization of the fact that the strong constraint in the 
ordinary DFT originates from the compatibility of Drinfel'd double for
Lie bialgebroids \cite{Mori:2019slw}.

\section{Relation to heterotic generalized geometry}
\label{sec:heterotic_gg}
We have discussed algebroids defined in the extended doubled space
$\mathcal{M}_{2D+n}$ in the gauged DFT.
These structures have deep connection to the generalized geometry
\cite{Hitchin:2004ut, Gualtieri} of heterotic string theories.
Indeed, it is known that upon the strong constraint, for ordinary (ungauged) DFT
case, the tangent bundle of the $2D$-dimensional doubled space
$\mathcal{M}_{2D}$ is identified with the generalized tangent bundle
through the so-called the natural isomorphism \cite{Freidel:2017yuv, Freidel:2018tkj}
;
\begin{align}
T \mathcal{M}_{2D} \simeq TM_D \oplus T^*M_D
\end{align}
where $M_D$ is the $D$-dimensional physical spacetime defined by the
strong constraint.
In this section, we discuss the relation between the algebroids on the
extended doubled space $\mathcal{M}_{2D+n}$ and the heterotic
generalized geometry.

Extended Courant algebroids that accompany gauge structures 
have been studied in various contexts.
In particular, in order to introduce non-Abelian gauge groups, 
the ordinary generalized tangent bundle is supplemented by the gauge sector.
This is obtained, for example,
by the generalized reduction of higher dimensional
Courant algebroid \cite{Bursztyn:2005}. 
Mathematically, this is a transitive Courant
algebroid \cite{Severa:2017oew} which is 
analyzed in the context of (exceptional)
generalized geometry \cite{Baraglia:2011}.
A related concept is the $B_n$-generalized geometry \cite{Rubio:2013} for which the ordinary
generalized tangent bundle 
$\mathbb{T} M_D = T M_D \oplus T^* M_D$
is extended to
$T M_D \oplus \mathrm{ad} P_G \oplus T^* M_D$.
Here $\mathrm{ad} P_G$ is 
the adjoint bundle of a principal $G$-bundle $P_G$ over $M_D$ 
for a gauge group $G$.
Applications of these extended Courant algebroids to heterotic theories
are discussed in \cite{Garcia-Fernandez:2014, Baraglia:2013}.
The generalized reduction in the mathematical literature has been
anticipated in the physical viewpoints \cite{Duff:1985cm, Duff:1986ya}.

As in the case of the (ungauged) $O(D,D)$ DFT, we solve the physical conditions
by all the quantities that depend only on $x^{\mu}$.
This implies 
$\tilde{\del}^{\mu} = \bar{\del}^{\alpha} = 0$ and hence 
$\tilde{\dop} = \bar{\dop} = [\cdot,\cdot]_{\tilde{L}} = [\cdot,
\cdot]_{\bar{L}} = 0$ and 
$\tilde{\mathcal{L}}_{\xi} = \bar{\mathcal{L}}_a = \bar{\mathfrak{L}}_X
= \bar{\mathfrak{L}}_{\xi} = \tilde{\mathfrak{L}}_a = 0$
in the twisted $C$-bracket \eqref{eq:twisted_C-bracket_geo}.
The physical spacetime is then the $D$-dimensional slice $M_D$ defined by
$\tilde{x}_{\mu} = \text{const.}$, $\bar{x}_{\alpha} = \text{const.}$
The terms in the gauge sector are 
\begin{align}
\frac{1}{2} 
\Big(
\mathfrak{L}_{a_1} a_2 - \mathfrak{L}_{a_2} a_1
\Big)
=& \ \frac{1}{2} \kappa^{\alpha \beta} 
\Big(
a_{1 \alpha} \del_{\mu} a_{2 \beta} - a_{2 \alpha} \del_{\mu} a_{1 \beta}
\Big) \tilde{\del}^{\mu},
\notag \\
\bar{\iop}_{a_2} \bar{\iop}_{a_1} F =& \
a_{2 \beta} a_{1 \gamma} F_{\alpha} {}^{\beta \gamma} \bar{\del}^{\alpha}.
\end{align}
The vector field $a = a_{\alpha} \bar{\del}^{\alpha} \in \Gamma (\bar{L})$ have the index $\alpha$
for the adjoint representation of the gauge group $G$ and we now identify
the basis $\bar{\del}^{\alpha}$ with the generator $T^{\alpha}$ of the
gauge group $G$.
Under these conditions, 
 we find that the twisted C-bracket \eqref{eq:twisted_C-bracket_geo} becomes
\begin{align}
[\Xi_1, \Xi_2]_F =& \  [X_1, X_2]_L + \mathcal{L}_{X_1} \xi_2 -
 \mathcal{L}_{X_2} \xi_1 - \frac{1}{2} \dop (\iop_{X_1} \xi_2 - \iop_{X_2} \xi_1)
\notag \\
& \ + \mathcal{L}_{X_1} a_2 - \mathcal{L}_{X_2} a_1 
+ \llbracket a_1, a_2 \rrbracket 
+ \mathrm{tr} \big[a_1 \dop a_2 - a_2 \dop a_1 \big],
\label{eq:heterotic_Courant_bracket}
\end{align}
where we have 
introduced the Lie algebra valued vector fields $a_{i} = a_{i \alpha}
T^{\alpha}$ and used the relation $\frac{1}{2} (\mathfrak{L}_{a_1} a_2 -
\mathfrak{L}_{a_2} a_1) = \mathrm{tr} \big[a_1 
\dop a_2 - a_2 \dop a_1 \big]$
and $\bar{\iop}_{a_2} \bar{\iop}_{a_1} F = 
\llbracket a_1, a_2 \rrbracket$ by 
using the relation 
$\llbracket T^{\alpha}, T^{\beta} \rrbracket
= - F_{\gamma} {}^{\alpha \beta} T^{\gamma}$
for the Lie algebra bracket $\llbracket \cdot, \cdot \rrbracket$ and the normalization 
$\mathrm{tr} \big[ T^{\alpha}
T^{\beta} \big] = \frac{1}{2} \kappa^{\alpha \beta}$.
The bracket \eqref{eq:heterotic_Courant_bracket} is nothing but the
heterotic Courant bracket on $TM_D \oplus \mathrm{ad} P_G \oplus T^*M_D$ \cite{Coimbra:2014qaa}.
This fact makes us to find a heterotic version of the natural
isomorphism \cite{Freidel:2017yuv, Freidel:2018tkj};
\begin{align}
T \mathcal{M}_{2D+n} 
\simeq 
TM_D \oplus \mathrm{ad} P_G \oplus T^*M_D.
\end{align}
Then in the physical spacetime $M_D$,
the tangent bundle of the extended doubled space $T\mathcal{M}_{2D+n}$ is identified
with the heterotic generalized tangent bundle.

\section{Conclusion and discussions} \label{sec:conclusion}
In this paper, we studied an extended doubled structure of algebroids defined by
the twisted C-bracket in the gauged DFT.

We first provided the geometrical framework for 
the twisted C-bracket in the $(2D+n)$-dimensional manifold
$\mathcal{M}_{2D+n}$ in which the
$2D$-dimensional para-Hermitian manifold is embedded.
This gives explicit expression of the twisted C-bracket in the
Drinfel'd double-like form that is different from the one for the C-bracket
in the ordinary DFT.
Using the geometric expression of the twisted C-bracket, we showed that
the bracket together with the appropriate bilinear form and the
anchor maps define a metric algebroid.
The fact that this is constructed out of three (twisted) Lie algebroids
is an analogue of the Drinfel'd double for Courant algebroids.
In order to implement the algebroid structure, 
we introduced in advance the notion of twisted Lie algebroids for which we find
the conditions on the flux $F_{\alpha} {}^{\beta \gamma}$.
They are just the Jacobi identity for $F_{\alpha} {}^{\beta \gamma}$ and the
gauge condition $F_{\alpha} {}^{\beta \gamma} \bar{\del}^{\alpha} * =
0$ in the gauged DFT.

We next studied the consistency conditions for Courant algebroids.
In addition to the requirement that $(L,\bar{L})$ form a Lie
bialgebroid \cite{Mackenzie}, 
we found extra conditions in the gauge sector.
These conditions are encoded, in the gauged DFT language, to the strong
constraint and the gauge condition.
The situation is similar to the Drinfel'd double for Courant algebroids
in the ungauged DFT but here it is extended to include the gauge sector.
When the conditions for the Courant algebroid are
imposed, we found that the twisted C-bracket reduces to the Courant bracket in
the heterotic theory.
We showed that this is associated with the heterotic version of the natural
isomorphism which allows us to identify the tangent bundle of the
extended doubled space $\mathcal{M}_{2D+n}$ with the heterotic
generalized tangent bundle.
This implies that the heterotic Courant algebroid is decomposed into
three pieces of algebroids.

In our analysis, it was shown that even for algebroids whose dimensions
are not even number, there could be doubled-like structures.
This implies that the gauged supergravities itself exhibit doubled-like
structure behind them 
\cite{Catal-Ozer:2023aea}.
The Drinfel'd double-like structures in gauged DFT will be useful
to analyze T-dualities among heterotic supergravity solutions and gauged
supergravities.
Relations to $L_{\infty}$-algebra are also interesting issues
\cite{Grewcoe:2020hyo, Hohm:2017pnh, Grewcoe:2020ren, Lescano:2021but}.
It would be interesting to find a global structure of the gauge
symmetry. For example, trying to the ``coquecigrue problem''
\cite{Severa, Severa2, Ikeda:2020lxz} {\it i.e.} finding the integrated,
group-like structure corresponding to the twisted C-bracket is an interesting direction.
We will come back to these issues in future studies.

%%%%%%%%%%%%%%%%%%%%%%%%%%%%%%%%%%%
\subsection*{Acknowledgments}
We are grateful to 
Jeong-Hyuck Park, Yuho Sakatani and Rui Sun for useful comments on 
algebroids and integrability during the CQUeST-APCTP Workshop ``Gravity
beyond Riemannian Paradigm''.
The authors would like to express their sincere gratitude to the reviewers
for their valuable feedback and insightful comments on this manuscript.
Their meticulous review process has greatly contributed to improving the
quality and accuracy of the paper.
The work of S.~S.  is supported in part by Grant-in-Aid for Scientific Research (C), JSPS KAKENHI
Grant Number JP20K03952.
The work of H.~M. is supported by Grant-in-Aid for JSPS Research Fellow,
JSPS KAKENHI Grant Numbers JP22J14419 and JP22KJ2651.

\begin{appendix}
\section{Doubled structures in $O(D,D)$ DFT} \label{Appendix:DFT}

The C-bracket in the $O(D,D)$ ungauged DFT is given by
\begin{align}
([e_{1}, e_{2}]_{\sf C})^{\hat{M}} = 
e_{1}^{\hat{K}} \del_{\hat{K}} e_{2}^{\hat{M}}
-
e_{2}^{\hat{K}} \del_{\hat{K}} e_{1}^{\hat{M}}
-
\frac{1}{2} \eta^{\hat{M}\hat{N}} \eta_{\hat{K}\hat{L}} 
\Big(
e_{1}^{\hat{K}} \del_{\hat{N}} e_{2}^{\hat{L}} - e_{2}^{\hat{K}} \del_{\hat{N}} e_{1}^{\hat{L}}
\Big),
\label{eq:ODD_C-bracket}
\end{align}
where $\hat{M},\hat{N}, \ldots = 1, \ldots, 2D$ and 
\begin{align}
\eta_{\hat{M}\hat{N}} = 
\left(
\begin{array}{cc}
0 & 1_D  \\
1_D & 0
\end{array}
\right),
\qquad
\eta^{\hat{M}\hat{N}}
=
\left(
\begin{array}{cc}
0 & 1_D  \\
1_D & 0
\end{array}
\right)
\end{align}
are the $O(D,D)$ invariant metric and its inverse.
The doubled vector $e^{\hat{M}}$ is decomposed as $e^{\hat{M}} = (X^{\mu}, \xi_{\mu})$
and the geometric expression of the C-bracket \eqref{eq:ODD_C-bracket}
is given by 
\begin{align}
[e_{1}, e_{2}]_{\sf C} =& \ 
[X_{1}, X_{2}]_{\hat{L}} + \mathcal{\hat{L}}_{\xi_{1}} X_{2} -
 \mathcal{\hat{L}}_{\xi_{2}} X_{1} - \frac{1}{2} \dop (\iop_{X_{1}} \xi_{2} - \iop_{X_{2}} \xi_{1})
\notag \\
& \ + [\xi_{1}, \xi_{2}]_{\tilde{\hat{L}}} + \mathcal{\hat{L}}_{X_{1}} \xi_{2} -
 \mathcal{\hat{L}}_{X_{2}} \xi_{1} - \frac{1}{2} \tilde{\dop} (\tilde{\iop}_{\xi_{1}} X_{2}
 - \tilde{\iop}_{\xi_{2}} X_{1}).
\label{eq:C-bracket_DFT}
\end{align}

\section{Detailed expressions of $I'_1$, $I'_2$, $I'_3$}\label{Appendix:AxiomC3}
The explicit form of the $\Gamma(\tilde{L})$ part $I'_1$ in
$\mathrm{Jac}_{F} (\Xi_{1}, \Xi_{2}, \Xi_{3})$ is given by
\footnotesize
\begin{align}
	I'_{1}
	&=
	[[\xi_{1}, \xi_{2}]_{\tilde{L}}, \xi_{3}]_{\tilde{L}}
	+
	\mathcal{L}_{[X_{1}, X_{2}]_{L}} \xi_{3}
	-
	\mathcal{L}_{X_{3}} [\xi_{1}, \xi_{2}]_{\tilde{L}}
	+
	\frac{1}{2} \dop \tilde{\iop}_{\xi_{3}} [X_{1}, X_{2} ]_{L}
	-
	\frac{1}{2} \dop \iop_{[\xi_{1}, \xi_{2}]_{\tilde{L}}} X_{3}
	\notag\\
	&\quad
	+
	\Bigl(
	[\mathcal{L}_{X_{1}} \xi_{2}, \xi_{3}]_{\tilde{L}}
	+
	\frac{1}{2} [\dop \tilde{\iop}_{\xi_{1}} X_{2}, \xi_{3}]_{\tilde{L}}
	+
	\mathcal{L}_{\tilde{\mathcal{L}}_{\xi_{1}} X_{2}} \xi_{3}
	-
	\frac{1}{2} \mathcal{L}_{\tilde{\dop} \tilde{\iop}_{\xi_{1}} X_{2}} \xi_{3}
	-
	\mathcal{L}_{X_{3}} \mathcal{L}_{X_{1}} \xi_{2}
	\notag\\
	&\qquad
	-
	\frac{1}{2} \mathcal{L}_{X_{3}} \dop \tilde{\iop}_{\xi_{1}} X_{2}
	+
	\frac{1}{2} \dop \tilde{\iop}_{\xi_{3}} \tilde{\mathcal{L}}_{\xi_{1}} X_{2}
	-
	\frac{1}{4} \dop \tilde{\iop}_{\xi_{3}} \tilde{\dop} \tilde{\iop}_{\xi_{1}} X_{2}
	-
	\frac{1}{2} \dop \tilde{\iop}_{\mathcal{L}_{X_{1}} \xi_{2}} X_{3}
	-
	\frac{1}{4} \dop \tilde{\iop}_{(\dop \tilde{\iop}_{\xi_{1}}X_{2})} X_{3} 
	-
	(1 \leftrightarrow 2)
	\Bigr)
	\notag\\
%%%%%
	&\quad
	+
	\frac{1}{2} \bar{\mathcal{L}}_{[a_{1}, a_{2}]_{\bar{L}}} \xi_{3}
	-
	\bar{\mathcal{L}}_{a_{3}} [\xi_{1}, \xi_{2}]_{\tilde{L}}
	+
	\frac{1}{4} \mathfrak{L}_{[a_{1}, a_{2}]_{\bar{L}}} a_{3}
	- 
	\frac{1}{4} \mathfrak{L}_{a_{3}} [a_{1}, a_{2}]_{\bar{L}}
	\notag\\
	&\quad
	+
	\Bigl(
	[\bar{\mathcal{L}}_{a_{1}} \xi_{2}, \xi_{3}]_{\tilde{L}}
	+
	\frac{1}{2}
	[\mathfrak{L}_{a_{1}} a_{2}, \xi_{3}]_{\tilde{L}}
	+ 
	\mathcal{L}_{\bar{\mathcal{L}}_{a_{1}} X_{2}} \xi_{3}
	+ 
	\frac{1}{2} \mathcal{L}_{\tilde{\mathfrak{L}}_{a_{1}} a_{2}} \xi_{3} 
	-
	\mathcal{L}_{X_{3}} \bar{\mathcal{L}}_{a_{1}} \xi_{2}
	-
	\frac{1}{2} \mathcal{L}_{X_{3}} \mathfrak{L}_{a_{1}} a_{2}
	\notag\\
	&\qquad
	+
	\frac{1}{2} \bar{\mathcal{L}}_{ \bar{\mathcal{L}}_{a_{1}} a_{2}} \xi_{3}
	+
	\bar{\mathcal{L}}_{\mathcal{L}_{X_{1}} a_{2}} \xi_{3}
	+
	\bar{\mathcal{L}}_{\tilde{\mathcal{L}}_{\xi_{1}} a_{2}} \xi_{3}
	+
	\frac{1}{2} \bar{\mathcal{L}}_{\bar{\mathfrak{L}}_{X_{1}} \xi_{2}} \xi_{3}
	+
	\frac{1}{2} \bar{\mathcal{L}}_{\bar{\mathfrak{L}}_{\xi_{1}} X_{2}} \xi_{3}
	-
	\bar{\mathcal{L}}_{a_{3}} \mathcal{L}_{X_{1}} \xi_{2}
	\notag\\
	&\qquad
	+
	\frac{1}{2} \bar{\mathcal{L}}_{a_{3}} \dop \iop_{X_{1}} \xi_{2}
	- 
	\bar{\mathcal{L}}_{a_{3}} \bar{\mathcal{L}}_{a_{1}} \xi_{2}
	- 
	\frac{1}{2} \bar{\mathcal{L}}_{a_{3}} \mathfrak{L}_{a_{1}} a_{2}
	+
	\frac{1}{4} \mathfrak{L}_{\bar{\mathcal{L}}_{a_{1}} a_{2}} a_{3}
	+
	\frac{1}{2} \mathfrak{L}_{\mathcal{L}_{X_{1}} a_{2}} a_{3}
	+
	\frac{1}{2} \mathfrak{L}_{\tilde{\mathcal{L}}_{\xi_{1}} a_{2}} a_{3}
	\notag\\
	&\qquad
	+
	\frac{1}{4} \mathfrak{L}_{\bar{\mathfrak{L}}_{X_{1}} \xi_{2}} a_{3}
	+
	\frac{1}{4} \mathfrak{L}_{\bar{\mathfrak{L}}_{\xi_{1}} X_{2}} a_{3}
	- 
	\frac{1}{4} \mathfrak{L}_{a_{3}} \bar{\mathcal{L}}_{a_{1}} a_{2}
	- 
	\frac{1}{2} \mathfrak{L}_{a_{3}} \mathcal{L}_{X_{1}} a_{2}
	- 
	\frac{1}{2} \mathfrak{L}_{a_{3}} \tilde{\mathcal{L}}_{\xi_{1}} a_{2}
	- 
	\frac{1}{4} \mathfrak{L}_{a_{3}} \bar{\mathfrak{L}}_{X_{1}} \xi_{2}
	\notag\\
	&\qquad
	- 
	\frac{1}{4} \mathfrak{L}_{a_{3}} \bar{\mathfrak{L}}_{\xi_{1}} X_{2}
	-
	\frac{1}{2} \dop \iop_{(\mathcal{L}_{a_{1}} X_{2})} \xi_{3}
	-
	\frac{1}{4} \dop \iop_{(\tilde{\mathfrak{L}}_{a_{1}} a_{2})} \xi_{3}
	+
	\frac{1}{2} \iop_{X_{3}}  \bar{\mathcal{L}}_{a_{1}} \xi_{2}
	+
	\frac{1}{4} \iop_{X_{3}}  \mathfrak{L}_{a_{1}} a_{2}
	-
	(1 \leftrightarrow 2)
	\Bigr)
	\notag\\
	&\quad
	+ 
	[\bold{i}_{\Xi_{2}} \bold{i}_{\Xi_{1}}F_{\mu} \tilde{\partial}^{\mu}, \xi_{3}]_{\tilde{L}}
	+ 
	\mathcal{L}_{(\bold{i}_{\Xi_{2}} \bold{i}_{\Xi_{1}}F^{\mu} \partial_{\mu})} \xi_{3}
	-
	\mathcal{L}_{X_{3}} (\bold{i}_{\Xi_{2}} \bold{i}_{\Xi_{1}}F_{\mu} \tilde{\partial}^{\mu})
	+ 
	\bar{\mathcal{L}}_{(\bold{i}_{\Xi_{2}} \bold{i}_{\Xi_{1}}F_{\alpha} \bar{\partial}^{\alpha})} \xi_{3}
	- 
	\bar{\mathcal{L}}_{a_{3}} (\bold{i}_{\Xi_{2}} \bold{i}_{\Xi_{1}}F_{\mu} \tilde{\partial}^{\mu})
	\notag\\
	&\quad
	+
	\frac{1}{2} \mathfrak{L}_{(\bold{i}_{\Xi_{2}} \bold{i}_{\Xi_{1}}F_{\alpha} \bar{\partial}^{\alpha})} a_{3}
	- 
	\frac{1}{2} \mathfrak{L}_{a_{3}} (\bold{i}_{\Xi_{2}} \bold{i}_{\Xi_{1}}F_{\alpha} \bar{\partial}^{\alpha})
	-
	\frac{1}{2} \dop \iop_{(\bold{i}_{\Xi_{2}} \bold{i}_{\Xi_{1}}F^{\mu} \partial_{\mu})} \xi_{3}
	+
	\frac{1}{2} \iop_{X_{3}} \bold{i}_{\Xi_{2}} \bold{i}_{\Xi_{1}}F
	+
	\bold{i}_{\Xi_{3}}\bold{i}_{[\Xi_{1}, \Xi_{2} ]_{F}} F_{\mu} \tilde{\partial}^{\mu}
	\notag\\
	&\quad
	+ \mathrm{c.p.}
	\label{eq:I1prime}
\end{align}
\normalsize
The first to the third lines in the right-hand side of
\eqref{eq:I1prime} are common with the $O(D,D)$ case. 
They are equal to $I_{1}$ in \cite{Mori:2019slw}.
The $\Gamma(L)$ part $I'_{2}$ in $\mathrm{Jac}_{F} (\Xi_{1}, \Xi_{2},
\Xi_{3})$ is the same.

The $\Gamma(\bar{L})$ part $I'_3$ in $\mathrm{Jac}_{F} (\Xi_{1},
\Xi_{2}, \Xi_{3})$ is given by 
\footnotesize
\begin{align}
	I'_{3}
	&=
	\frac{1}{4} [[a_{1}, a_{2}]_{\bar{L}}, a_{3}]_{\bar{L}}
	+
	\frac{1}{4} \bar{\mathcal{L}}_{[a_{1}, a_{2}]_{\bar{L}}} a_{3}
	-
	\frac{1}{4} \bar{\mathcal{L}}_{a_{3}} [a_{1}, a_{2}]_{\bar{L}}
	+
	\mathcal{L}_{[X_{1}, X_{2}]_{L}} a_{3}
	-
	\frac{1}{2} \mathcal{L}_{X_{3}} [a_{1}, a_{2}]_{\bar{L}}
	\notag\\
	&\quad
	+
	\tilde{\mathcal{L}}_{[\xi_{1}, \xi_{2}]_{\tilde{L}}} a_{3}
	-
	\frac{1}{2} \tilde{\mathcal{L}}_{\xi_{3}} [a_{1}, a_{2}]_{\bar{L}}
	+
	\frac{1}{2} \bar{\mathfrak{L}}_{[X_{1}, X_{2}]_{L}} \xi_{3}
	- 
	\frac{1}{2} \bar{\mathfrak{L}}_{X_{3}} [\xi_{1}, \xi_{2}]_{\tilde{L}}
	+
	\frac{1}{2} \bar{\mathfrak{L}}_{[X_{1}, X_{2}]_{L} } \xi_{3}
	-
	\frac{1}{2} \bar{\mathfrak{L}}_{\xi_{3}} [X_{1}, X_{2}]_{L} 
	\notag\\
%%%%%
	&\quad
	+
	\biggl(
	\frac{1}{4} [\bar{\mathcal{L}}_{a_{1}} a_{2}, a_{3}]_{\bar{L}}
	+
	\frac{1}{2} [\mathcal{L}_{X_{1}} a_{2} , a_{3}]_{\bar{L}}
	+
	\frac{1}{2} [\tilde{\mathcal{L}}_{\xi_{1}} a_{2}, a_{3}]_{\bar{L}}
	+
	\frac{1}{4} [\bar{\mathfrak{L}}_{X_{1}} \xi_{2}, a_{3}]_{\bar{L}}
	+
	\frac{1}{4} [\bar{\mathfrak{L}}_{\xi_{1}} X_{2}, a_{3}]_{\bar{L}}
	\notag\\
%%%%
	&\qquad
	+
	\frac{1}{4} \bar{\mathcal{L}}_{\bar{\mathcal{L}}_{a_{1}} a_{2}} a_{3}
	+
	\frac{1}{2} \bar{\mathcal{L}}_{\mathcal{L}_{X_{1}} a_{2}} a_{3}
	+
	\frac{1}{2} \bar{\mathcal{L}}_{\tilde{\mathcal{L}}_{\xi_{1}} a_{2}} a_{3}
	+
	\frac{1}{4} \bar{\mathcal{L}}_{\bar{\mathfrak{L}}_{X_{1}} \xi_{2}} a_{3}
	+
	\frac{1}{4} \bar{\mathcal{L}}_{\bar{\mathfrak{L}}_{\xi_{1}} X_{2}} a_{3}
	\notag\\
%%%%
	&\qquad
	-
	\frac{1}{4} \bar{\mathcal{L}}_{a_{3}} \bar{\mathcal{L}}_{a_{1}} a_{2}
	-
	\frac{1}{2} \bar{\mathcal{L}}_{a_{3}} \mathcal{L}_{X_{1}} a_{2}
	-
	\frac{1}{2} \bar{\mathcal{L}}_{a_{3}} \tilde{\mathcal{L}}_{\xi_{1}} a_{2}
	-
	\frac{1}{4} \bar{\mathcal{L}}_{a_{3}} \bar{\mathfrak{L}}_{X_{1}} \xi_{2}
	-
	\frac{1}{4} \bar{\mathcal{L}}_{a_{3}} \bar{\mathfrak{L}}_{\xi_{1}} X_{2}
%%%%%
	+
	\mathcal{L}_{\tilde{\mathcal{L}}_{\xi_{1}} X_{2}} a_{3}
\notag\\
&\qquad
	+ 
	\frac{1}{2} \mathcal{L}_{\tilde{\dop} \iop_{X_{1}} \xi_{2}} a_{3}
	+ 
	\mathcal{L}_{\bar{\mathcal{L}}_{a_{1}} X_{2}} a_{3} 
	+ 
	\frac{1}{2} \mathcal{L}_{\tilde{\mathfrak{L}}_{a_{1}} a_{2}} a_{3}
%%%%%
	-
	\frac{1}{2} \mathcal{L}_{X_{3}} \bar{\mathcal{L}}_{a_{1}} a_{2}
	-
	\mathcal{L}_{X_{3}} \mathcal{L}_{X_{1}} a_{2}
	-
	\mathcal{L}_{X_{3}} \tilde{\mathcal{L}}_{\xi_{1}} a_{2}
	\notag\\
	&\qquad
	-
	\frac{1}{2} \mathcal{L}_{X_{3}} \bar{\mathfrak{L}}_{X_{1}} \xi_{2}
	-
	\frac{1}{2} \mathcal{L}_{X_{3}} \bar{\mathfrak{L}}_{\xi_{1}} X_{2}
%%%%
	+ 
	\tilde{\mathcal{L}}_{\mathcal{L}_{X_{1}} \xi_{2}} a_{3}
	-
	\frac{1}{2} \tilde{\mathcal{L}}_{ \dop \iop_{X_{1}} \xi_{2}} a_{3}
	+
	\tilde{\mathcal{L}}_{\bar{\mathcal{L}}_{a_{1}} \xi_{2}} a_{3}
	+
	\frac{1}{2} \tilde{\mathcal{L}}_{\mathfrak{L}_{a_{1}} a_{2}} a_{3}
	\notag\\
%%%%
	&\qquad
	-
	\frac{1}{2} \tilde{\mathcal{L}}_{\xi_{3}} \bar{\mathcal{L}}_{a_{1}} a_{2}
	-
	\tilde{\mathcal{L}}_{\xi_{3}} \mathcal{L}_{X_{1}} a_{2}
	-
	\tilde{\mathcal{L}}_{\xi_{3}} \tilde{\mathcal{L}}_{\xi_{1}} a_{2}
	-
	\frac{1}{2} \tilde{\mathcal{L}}_{\xi_{3}} \bar{\mathfrak{L}}_{X_{1}} \xi_{2}
	-
	\frac{1}{2} \tilde{\mathcal{L}}_{\xi_{3}} \bar{\mathfrak{L}}_{\xi_{1}} X_{2}
%%%%%
	+ 
	\frac{1}{2} \bar{\mathfrak{L}}_{\tilde{\mathcal{L}}_{\xi_{1}} X_{2}} \xi_{3}
	\notag\\
	&\qquad
	+ 
	\frac{1}{4} \bar{\mathfrak{L}}_{\tilde{\dop} \iop_{X_{1}} \xi_{2}} \xi_{3}
	+ 
	\frac{1}{2} \bar{\mathfrak{L}}_{\bar{\mathcal{L}}_{a_{1}} X_{2}} \xi_{3}
	+ 
	\frac{1}{4} \bar{\mathfrak{L}}_{\tilde{\mathfrak{L}}_{a_{1}} a_{2}} \xi_{3}
%%%%
	- 
	\frac{1}{2} \bar{\mathfrak{L}}_{X_{3}} \mathcal{L}_{X_{1}} \xi_{2}
	+
	\frac{1}{4} \bar{\mathfrak{L}}_{X_{3}} \dop \iop_{X_{1}} \xi_{2}
	- 
	\frac{1}{2} \bar{\mathfrak{L}}_{X_{3}} \bar{\mathcal{L}}_{a_{1}} \xi_{2}
	\notag\\
	&\qquad
	-
	\frac{1}{4} \bar{\mathfrak{L}}_{X_{3}} \mathfrak{L}_{a_{1}} a_{2}
%%%%
	+ 
	\frac{1}{2} \bar{\mathfrak{L}}_{\tilde{\mathcal{L}}_{\xi_{1}} X_{2}} \xi_{3}
	+ 
	\frac{1}{4} \bar{\mathfrak{L}}_{\tilde{\dop} \iop_{X_{1}} \xi_{2}} \xi_{3}
	+ 
	\frac{1}{2} \bar{\mathfrak{L}}_{\bar{\mathcal{L}}_{a_{1}} X_{2}} \xi_{3}
	+ 
	\frac{1}{4} \bar{\mathfrak{L}}_{\tilde{\mathfrak{L}}_{a_{1}} a_{2}} \xi_{3}
	\notag\\
%%%%
	&\qquad 
	-
	\frac{1}{2} \bar{\mathfrak{L}}_{\xi_{3}} \tilde{\mathcal{L}}_{\xi_{1}} X_{2}
	-
	\frac{1}{4} \bar{\mathfrak{L}}_{\xi_{3}} \tilde{\dop} \iop_{X_{1}} \xi_{2} 
	-
	\frac{1}{2} \bar{\mathfrak{L}}_{\xi_{3}} \bar{\mathcal{L}}_{a_{1}} X_{2}
	-
	\frac{1}{4} \bar{\mathfrak{L}}_{\xi_{3}} \tilde{\mathfrak{L}}_{a_{1}} a_{2} 
	-
	(1\leftrightarrow2)
	\biggr)
	\notag\\
%%%%%
	&\quad 
	+ 
	\frac{1}{2} [\bold{i}_{\Xi_{2}} \bold{i}_{\Xi_{1}}F_{\alpha} \bar{\partial}^{\alpha}, a_{3}]_{\bar{L}}
	+ 
	\frac{1}{2} \bar{\mathcal{L}}_{(\bold{i}_{\Xi_{2}} \bold{i}_{\Xi_{1}}F_{\alpha} \bar{\partial}^{\alpha})} a_{3}
	-
	\frac{1}{2} \bar{\mathcal{L}}_{a_{3}} (\bold{i}_{\Xi_{2}} \bold{i}_{\Xi_{1}}F_{\alpha} \bar{\partial}^{\alpha})
	+ 
	\mathcal{L}_{(\bold{i}_{\Xi_{2}} \bold{i}_{\Xi_{1}}F^{\mu} \partial_{\mu})} a_{3}
	-
	\mathcal{L}_{X_{3}} (\bold{i}_{\Xi_{2}} \bold{i}_{\Xi_{1}}F_{\alpha} \bar{\partial}^{\alpha})
	\notag\\
	&\quad 
	+ 
	\tilde{\mathcal{L}}_{(\bold{i}_{\Xi_{2}} \bold{i}_{\Xi_{1}}F_{\mu} \tilde{\partial}^{\mu})} a_{3}
	-
	\tilde{\mathcal{L}}_{\xi_{3}} (\bold{i}_{\Xi_{2}} \bold{i}_{\Xi_{1}}F_{\alpha} \bar{\partial}^{\alpha})
	+
	\frac{1}{2} \bar{\mathfrak{L}}_{(\bold{i}_{\Xi_{2}} \bold{i}_{\Xi_{1}}F^{\mu} \partial_{\mu})} \xi_{3}
	- 
	\frac{1}{2} \bar{\mathfrak{L}}_{X_{3}} (\bold{i}_{\Xi_{2}} \bold{i}_{\Xi_{1}}F_{\mu} \tilde{\partial}^{\mu})
	\notag\\
	&\quad 
	+
	\frac{1}{2} \bar{\mathfrak{L}}_{(\bold{i}_{\Xi_{2}} \bold{i}_{\Xi_{1}}F^{\mu} \partial_{\mu})} \xi_{3}
	-
	\frac{1}{2} \bar{\mathfrak{L}}_{\xi_{3}} (\bold{i}_{\Xi_{2}} \bold{i}_{\Xi_{1}}F^{\mu} \partial_{\mu})
	+
	\bold{i}_{\Xi_{3}} \bold{i}_{[\Xi_{1}, \Xi_{2} ]_{F}} F_{\alpha} \bar{\partial}^{\alpha}
	+ \mathrm{c.p.}
	\label{eq:I3prime}
\end{align}
\normalsize

\end{appendix}

%%%%%%%%%%%%%%%%%%%%%%%%%%%%%%%%%%%%%%%%%%%%%%%%%

\end{document}